\documentclass[manuscript]{aastex}

\usepackage{lscape}
\usepackage{morefloats}
\usepackage{lineno}

\slugcomment{Not to appear in Nonlearned J., 45.}

\shorttitle{Traveling foreshocks}
\shortauthors{Kajdi\v{c} et al.}

\begin{document}

\title{Traveling foreshocks and transient foreshock phenomena}


\author{P. Kajdi\v{c}\altaffilmark{1}}
\affil{Instituto de Geof\' isica, Universidad Nacional Aut\'onoma de M\'exico, Mexico City, Mexico}
\email{primoz@igeofisica.unam.mx}
\author{X. Blanco-Cano\altaffilmark{1}}
\author{N. Omidi\altaffilmark{2}}
\affil{Solana Scientific Inc., Solana Beach, California, USA}
\author{D. Rojas-Castillo\altaffilmark{3}}
\affil{Swedish Institute of Space Physics, Box 812, Kiruna SE-98128, Sweden}
\author{D. G. Sibeck\altaffilmark{4}}
\affil{Code 674, NASA GSFC, Greenbelt, Maryland, USA}
\author{L. Billingham\altaffilmark{6}}
\affil{British Geological Survey, Edinburgh, UK}

\begin{abstract}
We use the multi-spacecraft capabilities of the Cluster and THEMIS missions to show that two types of foreshock may be detected in spacecraft data. One is the global foreshock that appears upstream of the Earth's quasi-parallel bow-shock under steady or variable interplanetary magnetic field. Another type is a traveling foreshock that is bounded by two rotational discontinuities in the interplanetary magnetic field and propagates along the bow-shock. Foreshock compressional boundaries are found at the edges of both types of foreshock. We show that isolated foreshock cavities are a subset of the traveling foreshock that form when two bounding rotational discontinuities are so close that the ultra-low frequency waves do not develop in the region between them. We also report observations of a spontaneous hot flow anomaly inside a traveling foreshock. This means that other phenomena, such as foreshock cavitons, may also exist inside this type of foreshock. In the second part of this work we present statistical properties of phenomena related to the foreshock, namely foreshock cavities, cavitons, spontaneous hot flow anomalies and foreshock compressional boundaries. We show that spontaneous hot flow anomalies are the most depleted transient structures in terms of the B-field and plasma density inside them and that the foreshock compressional boundaries and foreshock cavities are closely related structures.
\end{abstract}


\section{Introduction}

As the solar wind (SW) flows away from the Sun, it encounters obstacles, such as planets and their magnetospheres. Close to them, the SW is decelerated, deflected and heated by the shock waves that stand in front of these obstacles. Due to their shapes, these shock waves are referred to as bow-shocks. They are collisionless in nature because the mean free path of ions is much larger than the bow-shock sizes. 

The most studied bow-shock is the one standing in front of our planet. On average, its subsolar point is located $\sim$13~R$_E$ sunward of the Earth, but this distance can vary between 10~R$_E$ and 20~R$_E$ \citep[e.g.][]{meziane14}. Its Alfv\'enic and magnetosonic Mach numbers (M$_A$ and M$_{ms}$, respectively)  typically range between 6$\leq$M$_A\leq$ 7 and 5$\leq$M$_{ms}\leq$6 \citep[][]{winterhalterkivelson88}. Due to such high Mach numbers, the bow-shock of Earth is supercritical, meaning that it dissipates most of the SW kinetic energy by reflecting a portion of incident SW ions \citep[e.g., ][ and references therein]{treuman09}. 

An important parameter that determines what is observed upstream of the Earth's bow-shock in terms of waves and particles is the angle between the upstream interplanetary magnetic field (IMF) and the shock normal, $\theta_{BN}$. 
Most of the foreshock phenomena are observed for $\theta_{BN}<45^\circ$. Thus we commonly refer to the portion of the bow-shock with $\theta_{BN}$ less (more) than 45$^\circ$ as quasi-parallel (quasi-perpendicular) shock.

Observations however show backstreaming ions for $\Theta_{BN}\leq$70$^\circ$ \citep[e.g., ][]{eastwood05}. These ions exhibit relatively cold distributions and propagate upstream along the IMF, hence they are called field-aligned ion beams  \citep[FAB; ][]{gosling78, gosling79, thomsen85, kis07, meziane13}. Their energies tend to be $\lesssim$10~keV. FABs interact with  the incoming SW particles and this can result in the growth of ultra-low frequency (ULF) waves \citep[e.g., ][]{gary93, dorfman17} with typical periods of $\sim$30~s. Since they need some time to grow, the ULF waves are not observed together with the FABs but rather together with the so called intermediate ion distributions \citep[][]{paschmann79}. Finally, as the ULF waves propagate through regions where suprathermal particles exhibit strong density gradients they steepen and thus gain a significant compressive component. Such waves are observed together with diffuse ion populations \citep[e.g., ][]{fuselier86, kis04, eastwood05}. The diffuse and intermediate ions exhibit energies up to several hundreds of KeV. FABs, intermediate and diffuse ions are commonly called suprathermal ions. The region upstream of Earth's bow-shock populated by ULF waves (suprathermal ions) is called the ULF wave (suprathermal ion) foreshock \citep[e.g., ][and references therein]{eastwood05}.

The ULF waves propagate sunwards in the SW frame of reference but are convected by the SW towards the bow-shock. As ULF waves approach the bow-shock, they steepen and can form shocklets \citep[e.g.][]{hopperussell81, hopperussell83, hadakennel87} and short-large amplitude magnetic structures \citep[SLAMS, e.g.][]{thomsen90, schwartzburgess91, schwartz92, mann94, lucek02}. The interaction of compressive and transverse ULF waves leads to the formation of foreshock cavitons \citep[][]{omidi07, bcano09, bcano11, kajdic11, kajdic13}. Cavitons convected by the SW generate spontaneous hot flow anomalies \citep[SHFAs, ][]{zhang13, omidi13b, omidi14} when they arrive to the bow-shock.

Another structure commonly observed at the edges of the foreshock is the foreshock compressional boundary \citep[FCB, ][]{omidi09, RojasCastillo13}. These structures separate either the pristine solar wind or the region populated by field-aligned ion beams from the region of the foreshock populated by compressive ULF waves and diffuse ions.

FCBs have been associated with foreshock cavities \citep[][]{schwartz06, billingham08, billingham11}. While the earlier works referred to foreshock cavities as isolated structures, \citet{billingham11} talk about boundary cavities that are found at the edges of the foreshock and were later referred to as FCBs. \citet{omidi13a} performed global hybrid simulations of planetary bow-shock, under varying upstream conditions. Specifically, the authors reproduced foreshock cavities by launching two consecutive IMF rotational discontinuities between which the IMF connected to the otherwise quasi-perpendicular bow-shock in such a way that the local $\theta_{BN}$ was less than 45$^\circ$. This lead to the development of foreshock-like regions upstream of a portion of the simulated bow-shock between the two IMF discontinuities, that were convected along the bow-shock surface. These regions were called by the authors foreshock cavities and also traveling foreshocks. FCBs formed at the edges of these regions.

In the first part of this work we use THEMIS and Cluster multi-spacecraft observations to perform case studies of foreshocks and foreshock cavities to confirm some of the predictions made by \citet{omidi13a}: we show that the spacecraft sometimes observe the global Earth's foreshock and sometimes a traveling foreshock. The global foreshock may be observed upstream of the quasi-parallel section of the Earth's bow-shock under either steady or variable IMF conditions. When the IMF changes its orientation, the foreshock changes its location with respect to the bow-shock. Two consecutive IMF rotations may cause the global foreshock to rock back and forth, resulting in a spacecraft initially located in the unperturbed solar wind to enter and then exit the foreshock.

We note here that traveling foreshocks should not be mistaken for another type of transient localized foreshocks that has recently been discovered by \citet[][]{Pfau-Kempf16}, which occur due to bow-shock perturbations caused by flux transfer events under stable solar wind and IMF conditions.

In a different scenario, an IMF flux tube is convected along the bow-shock. The spacecraft observes two IMF rotational discontinuities (RDs. Note: here we do not distinguish between rotational and tangential discontinuities). During the time between the RDs, the geometry of a portion of the bow-shock may change from quasi-perpendicular ($\theta_{BN}>$45$^\circ$) to quasi-parallel ($\theta_{BN}<$45$^\circ$), which leads to the formation of a region between the RDs that is populated by suprathermal particles and ULF fluctuations. As the two RDs propagate along the bow-shock, so does the perturbed region between them. We call such a region a traveling foreshock.

The only way to observationally distinguish between the back and forth motion of the global foreshock and the traveling foreshock is by using simultaneous observations of several spacecraft. In the first case the spacecraft observe the arrival of the foreshock at slightly different times in a certain sequence. If the spacecraft spatial configuration does not change, then the sequence in which they exit the foreshock is reversed. Such signatures in the spacecraft data are known as nested signatures \citep[e.g. ][]{burgess05}. On the other hand the sequence in which the spacecraft observe the traveling foreshock is the same as the sequence in which they exit it. The so-called convected signatures \citep[][]{burgess05} can be found in the spacecraft data in this case.

In the second part of this work we statistically compare observational properties of foreshock cavities, foreshock cavitons, foreshock compressional boundaries and spontaneous hot flow anomalies. We also compare their locations and the solar wind and IMF conditions under which they are observed.

This paper is organized as follows: in section 2.1 we present the instruments and data used in this study. In 2.2 we show multi-spacecraft observations of the global foreshock, the traveling foreshocks and foreshock cavities. In 2.3 we exhibit statistics of observational properties of several types of transient foreshock phenomena. In section 3 we discuss the results, and in section 4 we summarize our findings.

\section{Observations}
\subsection{Instruments and datasets}
We use multi-spacecraft data provided by the Cluster and THEMIS missions. 

The Cluster mission consists of four identical spacecraft that provide magnetic field and plasma measurements in the near-Earth environment. The spacecraft carry several instruments, including a Fluxgate Magnetometer \citep[FGM, ][]{balogh01} and the Cluster Ion Spectrometer \citep[CIS, ][]{reme01}. We use FGM magnetic field vectors and CIS-HIA solar wind ion moments with 0.2~s and 4~s time resolution, respectively.

The THEMIS mission consists of five spacecraft. Their Flux Gate Magnetometer \citep[][]{auster08} measures the background magnetic field with time resolution up to 64~Hz. Here we use data with 0.25~s resolution. The THEMIS ion and electron analyzers \citep[iESA and eESA,][]{mcfadden08} provide plasma moments and spectrograms with a spin (3~s) time resolution.

The data were accessed through the European Space Agency's Cluster Science Archive (http://www.cosmos.esa.int/web/csa) and through the ClWeb portal (http://clweb.irap.omp.eu) which is maintained by the 
Institut de Recherche en Astrophysique et Plan\'etologie (IRAP).

\subsection{Case studies}
\subsubsection{The global foreshock}
This section presents THEMIS observations of the global foreshock. THEMIS~A observed the foreshock on 7 August 2007 between 2:10~UT  and 2:44~UT. Figure~\ref{fig:20070807a} shows the data between 01:51~UT and 03:03~UT on the same day. The following quantities are displayed in the panels from top to bottom: a) magnetic field magnitude in units of nanoTesla (nT), b) magnetic field components in GSE coordinate system in units of nT, c) angle between the IMF and the Sun-Earth line in degrees, d) IMF clock angle in degrees, e) SW density in cm$^{-3}$, f) solar wind speed (black) and -V$_x$ component (red) in kms$^{-1}$, g) V$_y$ and V$_z$ components of SW velocity in kms$^{-1}$, h) SW temperature in eV, i) ion spectra with colors representing the logarithm of the particle energy flux (units eV/(cm$^-1\cdot$s$\cdot$str$\cdot$eV)). j), Morlet wavelet spectrum for B-magnitude, k) Morlet wavelet spectrum for Bx-component and l) B-magnitude and Bx component between 02:25-02:30~UT.

We can see that from 01:51~UT to 02:09~UT the IMF was relatively steady with only small rotations. The $\theta_{BX}$ (which is similar to $\theta_{BN}$ near the Sun-Earth line) displayed values between 60$^\circ$ and 90$^\circ$. During this time the THEMIS A spacecraft observed the pristine SW. At 02:09~UT (first vertical red line) the $\theta_{BX}$ starts diminishing until $\sim$02:16~UT (second vertical red line) when it reached values below 20$^\circ$. During this time interval the THEMIS A spacecraft entered the foreshock region. Several things point to that: the spacecraft became immersed in strong B-field fluctuations with amplitudes $\delta$B/B up to 0.5 and periods of several tens of seconds. These fluctuations contained a significant compressive component and are known as ULF waves. Intense fluctuations also appeared in the density and velocity panels. The bottom panel revealed the onset of suprathermal ion population (energies $\lesssim$30~keV) starting at $\sim$02:12~UT. 

Inside this foreshock the IMF and plasma parameters change with respect to the upstream solar wind: the average B-field magnitude decreases from $\sim$8.5~nT to $\sim$7.5~nT, the average plasma density from $\sim$4.3~cm$^{-3}$ to $\sim$3.8~cm$^{-3}$ and the average plasma velocity from $\sim$573~kms$^{-1}$ to $\sim$357~kms$^{-1}$. A detailed inspection of the ion spectrum in Figure~\ref{fig:20070807a} reveals that this decrease of plasma velocity is not due to the deceleration of the incident solar wind so it must be due to the contribution of suprathermal ions arriving from the Earth's bow-shock to the total plasma bulk velocity. We know this since the energy of the peak of the SW beam does not change. The latter primarily diminishes due to the Vx component, while the absolute values of the Vz component increase slightly (from $\sim$82~kms$^{-1}$ to $\sim$102~kms$^{-1}$). At 02:38~UT (third vertical red line) the $\theta_{BX}$ starts to increase again until 02:44~UT (fourth vertical red line). After that time the $\theta_{BX}$ values stay above 50$^\circ$ and the plasma and IMF parameters are steady. The suprathermal ions disappear at 02:43~UT. The spacecraft stays in the SW for the next few minutes. Before the end of the shown time interval at 03:03~UT, the spacecraft detects the suprathermal ions and ULF compressive fluctuations several more times but for shorter time intervals.

The two regions shaded in green in Figure~\ref{fig:20070807a} mark the intervals when a foreshock compressional boundary (FCB) is detected. These phenomena \citep[see for example][]{RojasCastillo13} are commonly observed at the edges of the foreshock and are characterized by correlated increments in B and N above the upstream SW values, followed by a drop below the upstream SW values. 

The first FCB is quite weak with only a small hump in B and N. The trailing FCB is much more prominent. The suprathermal ions appear and disappear just when the B and N inside the two FCBs reach their maximum values at 02:14~UT and 02:43~UT, respectively.

We examine the ULF wave properties more closely in Figure~\ref{fig:20070807a}, panels j and l. We can see that the ULF waves are compressive since both wavelet spectra exhibit similar power and since the B-magnitude in panel l) shows irregular ULF fluctuations. The frequencies of these waves are between 2$\times$10$^{-1}$Hz-8$\times$10$^{-2}$Hz corresponding to periods between $\sim$10~s-50~s.

In order to answer the question about which type of foreshock (global or traveling) THEMIS A observed, we perform a multi-spacecraft analysis of this event. As can be seen in Figure~\ref{fig:20070807b} the THEMIS spacecraft were in a string-of-pearls configuration along y$_{GSE}$ direction. At the time of the event they were positioned fairly near the Sun-Earth line. The leading spacecraft along their orbits was THEMIS B and the trailing THEMIS A, with the C, D, and E spacecraft located close together between THEMIS A an B. In Figures~\ref{fig:20070807c}a) and b) we show a closeup of B-field profiles of the leading and trailing parts of the foreshock interval presented in Figure~\ref{fig:20070807a}. Since the leading FCB was weak, we examine a structure observed just after it. It is shaded in orange in Figure~\ref{fig:20070807a}. The red trace in Figures~\ref{fig:20070807c}a) and b) corresponds to THEMIS A, the purple to THEMIS B, while the thin black traces, which can hardly be distinguished from each other, correspond to the C, D and E spacecraft. In Figure~\ref{fig:20070807c}a we see that THEMIS A detected the structure first (starting at $\sim$02:13:18~UT), while THEMIS B ($\sim$02:13:29~UT) was the last to observe it. Spacecraft C, D and E entered the structure roughly at 02:13:27~UT.

We can see in Figure~\ref{fig:20070807c}b that the sequence in which the spacecraft observed the trailing FCB, was reversed. THEMIS B detected this FCB starting at $\sim$02:41:15~UT, slightly before the C, D and E spacecraft {($\sim$02:41:17~UT) while THEMIS A was the last to detect it $\sim$02:41:24~UT. The B, C, D and E spacecraft observed a more extended FCB than the A spacecraft. This is probably because 1) each spacecraft crossed the structure at slightly different place and 2) they detected it at slightly different times during which the FCB could have evolved.

Figure~\ref{fig:sketch20070807} illustrates the situation in near-Earth interplanetary space around the times of arrival of the first (\ref{fig:sketch20070807}a) and the second (\ref{fig:sketch20070807}b) IMF rotation.  \citet[][]{fairfield71} models for bow-shock and magnetopause have been used here. In the Figure the X$_{GSE}$ axis points up while the Y$_{GSE}$ axis points right. In Figure~\ref{fig:sketch20070807}a we see that the initial IMF orientation is such that the foreshock is located on the left side of the Figure. At the first IMF rotation the foreshock becomes distorted, since the backstreaming ions follow the IMF lines that are connected to the bow-shock surface in such a way that they make a small angle with its normal (nominally $\theta_{BN}<$45$^\circ$). Upstream of the first rotation, the IMF is more radial and the foreshock shifts in the positive Y$_{GSE}$ direction. This means that at some point during the rotation spacecraft will enter the foreshock region and it will observe a FCB. Once the rotation reaches the bow-shock, we have an almost radial foreshock.
After some time a second IMF rotation arrives (Figure~\ref{fig:sketch20070807}b). Upstream of this rotation the $\theta_{Bx}$ angle increases again, which causes the foreshock to move back towards the left side of the Figure. Once this IMF rotation sweeps across the spacecraft and reaches the bow shock, the foreshock will be located the same way as it was before the arrival of the first IMF rotation. The spacecraft will move out of the foreshock and it will again observe a FCB. 

\subsubsection{The traveling foreshock}
\label{sec:TFSH}
Our next case study ocurred on 14 August 2007 between 20:56~UT and 21:13~UT (Figure~\ref{fig:20070814a}). The five THEMIS spacecraft were in a configuration similar to that in the previous case (see Figure~\ref{fig:20070814b}). Figure~\ref{fig:20070814a}a) presents the foreshock region similar to that in the previous case, but now it is bounded by two IMF rotations both of which occur on much shorter time scales (a few seconds) than the rotations in the previous case study ($\sim$seven minutes), hence we will call them rotational discontinuities (RD). They are marked with vertical red lines. Before the first RD the angle between the IMF and the Sun-Earth line ($\theta_{BX}$) was $\sim$70$^\circ$ and after the second RD it was above 80$^\circ$. During the time between the two RDs Bx oscillated between 0$^\circ$ and 90$^\circ$ with an average value around $\sim$40$^\circ$. As before, the two intervals shadowed in green mark the leading and trailing FCBs. We call the region between the two rotational discontinuities traveling foreshock for reasons that will become clear later. This region is populated by compressive ULF fluctuations and suprathermal ions.

Figure~\ref{fig:20070814c}a shows B-field magnitude profiles for the leading (panel a) and the trailing (b) FCBs. In panel a) we see that after about two minutes of steady B-field with magnitude of $\sim$5~nT observed by all spacecraft at slightly different times, the first to detect the leading FCB is THEMIS B (starting at $\sim$20:55:15~UT), followed by C, D and E $\sim$20:55:19~UT) spacecraft and THEMIS A is the last to detect it ($\sim$20:55:35~UT). The detection times are marked with vertical lines in the Figure. The same order is observed at the exit from the traveling foreshock, as it can be seen in Figure~\ref{fig:20070814c}b.
The fact that the order in which the spacecraft observed this foreshock is the same when they enter and when they exit it tells us that this foreshock swept across the spacecraft, hence we call it a traveling foreshock.

Figure~\ref{fig:sketch20070814}a illustrates the situation in this case. Again, \citet[][]{fairfield71} models for bow-shock and magnetopause have been used here. The purple color represents the global foreshock and the red color a flux tube with different orientation than the background IMF (blue lines). The black arrows show the orientation of the local bow-shock normal. The purple, black and red crosses represent spacecraft in a configuration similar to that in Figure~\ref{fig:20070814b}. The magnetic flux tubes are carried antisunward (downwards in the Figure) by the solar wind. Therefore the intersection of the red flux tube with the bow shock propagates along the Y$_{GSE}$ direction. The traveling foreshock also propagates in the same direction (indicated by the red arrow).  The local bow-shock geometry changes to quasi-parallel at places where the tube's field lines connect to the bow-shock, so a foreshock region forms upstream of this portion of the bow-shock. This is a traveling foreshock that propagates along the Y$_{GSE}$ axis due to the way in which B-field lines in it are oriented. Note that in this Figure the width of the flux tube is smaller than the width of the bow-shock, while in reality it can be larger.

We calculate the average orientation of the magnetic field inside the traveling foreshock (hence the orientation of the flux tube) in GSE coordinates to be (-0.59, 0.82, -0.20) while the solar wind velocity in it was $\sim$310~kms$^{-1}$, predominantly in the negative X$_{GSE}$ direction. The event lasted for 17~minutes in the spacecraft data. From this we can estimate the width of the flux tube to be 2.55$\cdot$10$^{5}$~km or about 40~R$_E$.

In Figure~\ref{fig:20070814a} we show the wavelet spectra of B (j) and Bx (k) and a five minute zoom on the B and -Bx (l). Again, we see that the ULF waveforms are highly irregular and that they exhibit a strong compressible component.

In the next section we show that the mechanism that is responsible for the formation of traveling foreshocks is basically the same as the mechanism for the formation of another structure, called isolated foreshock cavities \citep{schwartz06, billingham08}. We suggest that isolated foreshock cavities can be considered a subset of traveling foreshocks.

\subsubsection{Foreshock cavities}
This section presents two foreshock cavities. The first (Figure~\ref{fig:20051228a}) was observed by the Cluster quartet on 28 December 2005.
The event was first described by \citet[][]{billingham08}. It lasted for about a minute between 14:14~UT and 14:15~UT. It is bounded by two IMF RDs that are marked by red vertical lines. On the bottom panel we see that suprathermal ions are present during the time between the two IMF RDs. In Figure~\ref{fig:20051228a}b we show B-field magnitude profiles of the four spacecraft between 14:13~UT and 14:16~UT. The black, blue, green and red colors are for C1, C2, C3 and C4 spacecraft, respectively. The spacecraft entered into this cavity in the following sequence: C2, C1, C4, C3 and they exited it in the same order. This means that the cavity was convected past them. In this case the bounding rotations are close together and ULF waves are not observed between them.

Another foreshock cavity was observed on 14 August 2007 between 20:32~UT and 20:36~UT by the THEMIS-B spacecraft (Figure~\ref{fig:20051228c}a). This event is somewhat different from the previous case in the sense that the two IMF RDs (red vertical lines) are more separated and there are few ULF fluctuations that appear between them. Still, we classify this case study as a foreshock cavity as it resembles those published in the literature \citep[see for example Figure~3 of][]{sibeck02}.

The B-field magnitude data of the five THEMIS spacecraft are exhibited in Figure~\ref{fig:20051228c}b. The signature of the structure is different in the THEMIS A data, but we can still see that this is a convected structure since the order in which the spacecraft entered is the same as the order in which they exited it.

These two events share some similarities with traveling foreshocks: inside both of them we observe suprathermal ions, they are convected structures, both are delimited by IMF RDs and in the case of the longer lasting cavity observed on 14 August 2007 there are even compressive ULF fluctuations inside it. 
The difference between the two cases shown here and the traveling foreshock shown in Section~\ref{sec:TFSH} is that in the case of isolated cavities only a few or no ULF wave forms appear between the bounding IMF RDs, while in case of the traveling foreshock many compressive waves can be seen in the magnetic field and plasma data. Hence we conclude that the events that are commonly called foreshock cavities are a subset of traveling foreshocks.

\subsubsection{SHFA in a traveling foreshock}
Our last case study is shown in Figure~\ref{fig:20070809}. It was observed on 9 August 2007 between 19:42:30~UT and 19:53~UT by the THEMIS spacecraft. The detailed inspection of multi-spacecraft observations reveals that this is a traveling foreshock which is bounded by two B-field rotations, marked by two vertical red lines in Figure~\ref{fig:20070809}. The first rotation was particularly strong and it marks the onset of field-aligned suprathermal ions with energies up to $\sim$9~keV, which can be seen on the bottom panel. The magnetic field and density perturbations are small until about 19:48~UT. After that time there are compressive ULF fluctuations with $\delta$B/B$\sim$0.5 and the intensity of the suprathermal ions increases. The energy range of these ions extends to much higher energies indicating a diffuse population. At 19:53:04~UT there is a less prominent B-field rotation after which the suprathermal ions are still present although their intensity is much smaller and the B-field and density perturbations disappear. The spacecraft again entered the region of the foreshock populated by the field-aligned ion beams \citep[e.g.][]{eastwood05}. At this time there is also a FCB.

An interesting feature appearing on this Figure is a spontaneous hot flow anomaly (SHFA) centered at 19:50:58~UT (shaded in green in Figure~\ref{fig:20070809}). The SHFA exhibits typical signatures: B and N diminish at its center, but they are enhanced on its rims. There is an obvious increase of the proton temperature at the center (from 730~eV to 1340~eV). The absolute value of the $x$ component of the plasma velocity decreases from 339~kms$^{-1}$ to 192~kms$^{-1}$, the $z$ component changes from -65~kms$^{-1}$ to -264~kms$^{-1}$ and the total plasma velocity diminishes from 334~kms$^{-1}$ to 241~kms$^{-1}$. This feature is not associated with a tangential IMF discontinuity, hence we classify it as a spontaneous HFA, following the work of \citet{zhang13} and \citet{omidi13b, omidi14}. According to these authors the SHFAs occur when foreshock cavitons \citep[see][]{omidi07, bcano09, bcano11, kajdic11, kajdic13} interact with the Earth's bow-shock. 
This case study tells us that the foreshock transient structures, such as foreshock cavitons and SHFA can form inside traveling foreshocks.

\subsection{Statistical comparison of observational properties of upstream phenomena}
In the previous section we made several claims. For example, we pointed towards the relation between traveling foreshocks, FCBs and foreshock cavities. We also showed that transient phenomena, such as SHFAs can occur inside the traveling foreshocks. In this section we further strengthen our case by making use of statistical properties of these phenomena, namely isolated foreshock cavities, foreshock cavitons, spontaneous hot flow anomalies and foreshock compressional boundaries. The data for all phenomena except SHFAs was compiled from the already existing literature. The cavity statistics were published by \citet{billingham08} (over 200 events), caviton properties by \citet{kajdic13} (92 events) and those for the FCBs by \citet{RojasCastillo13} (36 events). To this we add statistics of 19 SHFAs found in the Cluster data between the years 2003 and 2011 (listed in Table~\ref{tab1}).

Histograms in Figure~\ref{fig:distributions} show relative changes in the (from left to right) magnetic field magnitude, plasma density and velocity and durations of (from top to bottom) SHFAs, foreshock cavitons, foreshock cavities and FCBs. $\Delta$ sign marks the difference between the ambient SW value and the minimum value inside the structure \citep[see ][ for details]{kajdic13, RojasCastillo13, billingham08}. In the case of FCBs it represents the difference between the maximum value inside the FCB and the upstream SW value. The upstream values were obtained by averaging the quantities during intervals adjacent to the events. The lengths of these intervals were typically of several tens of seconds up to a few minutes, although the extact lengths are different for each event. We can see that SHFAs are by far the most depleted structures with average $\Delta$B/B and $\Delta$N/N values of 0.9. In the case of the other three phenomena the average values of $\Delta$B/B and $\Delta$N/N are 0.5 for cavitons, 0.4 for cavities and 0.4 for FCBs and the spread in values is much larger. The velocity does not change inside the cavitons (which is one of the criteria to identify them), it changes slightly in the case of foreshock cavities and across FCBs, while the change is significant in the case of SHFAs. The average durations of cavitons and SHFAs are similar (about one minute), while they are longer ($\sim$107~seconds) in the case of foreshock cavities. All the described events last less than 200 seconds in the spacecraft data.

Figure~\ref{fig:dndb} shows a scatter plot of $\Delta$N/N versus $\Delta$B/B for the four phenomena. We can see that in the case of FCBs (purple plus signs) and foreshock cavitons (black asterisks) the two quantities are well correlated with correlation coefficients of 0.86 and 0.85, respectively. The correlation is less strong in the case of foreshock cavities (red diamonds, k = 0.63) and SHFAs (blue squares, k=0.30). We can see again that the SHFAs cluster at highest values and are hence the most depleted structures.

Next we look at locations of these phenomena in solar foreshock coordinates (SFC, Figure~\ref{fig:sfcsystem}). These coordinates were first introduced by \citet{greenstadtbaum86} in their study of the location of the ULF compressional waves in the Earth's foreshock. \citet{mezianeduston98} used these coordinates to describe the observed locations of the intermediate ion boundary. \citet{billingham08} used them for foreshock cavities while \citet{kajdic13} compared solar foreshock coordinates of foreshock cavitons to those of intermediate ions and ULF waves.
In order to calculate SFC we must first determine the cross section of a model bow-shock with a plane defined by the $x$ axis and IMF direction. On this plane we define a set of rectangular coordinates (x,$\eta$). The SFC consist of another set of coordinates (X$_f$, D$_{BT}$). X$_f$ is parallel to the Sun-Earth line and measures the distance between the observed structure and the tangential IMF line. D$_{BT}$ measures the distance along this line between its intersection with the bow-shock and the point with the same $\eta$ coordinate as the observed structure.

To calculate SFC of our events, we model the bow-shock shape as a hyperboloid and we use the solar wind dynamic pressure in order to scale the shock. We do this by first measuring solar wind properties during time intervals when the spacecraft were in the solar wind but were close to times when the structures (cavitons, etc.) were observed. We then calculate the dynamic pressure and follow the procedure described in \citet{jelinek2012} in order to obtain the stand-off distances of the bow-shock. Next we calculate the ratio between each calculated stand off distance and the stand-off distance of the nominal bow-shock model used by \citet{greenstadt72} and \citet{greenstadtbaum86}. The coordinates are calculated as explained in \citet{greenstadtbaum86}. 

The locations of the structures in SFC coordinates are presented in Figure\ref{fig:sfc}a.  In this Figure the horizontal green line represents the tangent line. The dashed blue line is a fit to the ULF wave boundary by \citet{greenstadtbaum86}, while the yellow dashed line represents a fit to ion intermediate boundary from \citet{mezianeduston98}. The black continuous line is a fit to caviton locations from \citet{kajdic13}. Black asterisks represent locations of foreshock cavitons, blue triangles of the SHFAs, red diamonds of the foreshock cavities and purple stars of the FCBs.

\citet{kajdic13}, \citet{billingham08} and \citet{mezianeduston98} all used a single bow-shock model for all their events. It can be seen in the Figure that the locations of the structures (for example foreshock cavitons) calculated by us are very different from those in the past literature. Our approach with the bow-shock scaled with the solar wind dynamic pressure is more accurate. One example to sustain this claim is that in Figure~\ref{fig:sfc}a there are no events outside the tangent line, i.e., in the solar wind that is not magnetically connected to the bow-shock, while this was the case when a single bow-shock model was used.

The first phenomena observed downstream of the tangent line are foreshock cavities (red diamonds). Foreshock cavitons (black asterisks) lie further downstream as expected, since the cavitons are always surrounded by compressive ULF waves. 
FCBs (purple stars) occupy the same region as cavities. This is because FCBs can appear at the edges of the traveling foreshocks and these are related to cavitites. Finally, SHFAs (blue triangles) tend to be found downstream of the cavitons and closer to the bow-shock than the rest of the phenomena.

Figure~\ref{fig:sfc}b shows the distributions of the angles $\theta_{BN}$ of the portion of the model bow-shock which different phenomena were magnetically connected to. These angles were calculated by obtaining the bow-shock shape and size following the work of \citet{greenstadt72}. The vast majority of angles for SHFAs and foreshock cavitons are smaller than 50$^\circ$, as expected. There are a few outliers. A possible explanation for these events is that they occured inside the traveling foreshocks so that the IMF vector, needed to calculate the $\theta_{BN}$ were obtained upstream of these foreshocks.

Foreshock cavities show a broad distribution of $\theta_{BN}$ peaking between 40$^\circ$ and 60$^\circ$ while a more flat distribution is seen in case of FCBs.

Figure~\ref{fig:sfc}c shows the distance (along the IMF direction) of the events to the model bow-shock. Most events were observed at distances $\leq$12~R$_E$, although this may partially be due to the fact that they were all observed by the Cluster spacecraft, which, when located upstream of the Earth's bow-shock, tend to stay close to it. 

SHFAs were all observed at distances $\leq$6~R$_E$, which is expected, since they are suppose to form due to cavitons interacting with the bow-shock. The distances of several R$_E$ could be partially explained by the fact that SHFAs have finite sizes (the sizes of SHFAs may be similar to those of foreshock cavitons, which, as has been shown by \citet{kajdic13}, are in rare cases more than 8~R$_E$). Another point is that these distances are along the IMF direction so SHFAs can actually be located upstream of a portion of the bow-shock to which they do not seem to be magnetically connected, but is closer to them.

Foreshock cavitons were also mostly observed at distances $\leq$5~$R_E$ which is also expected since they are found in the regions containing compressive ULF waves. The distributions of distances of foreshock cavities peaks between $\sim$3~$R_E$ and $\sim$5~R$_E$, while that of FCBs is relatively flat between 0~R$_E$ and 7~R$_E$.

We put all known phenomena in context in Figure~\ref{fig:sfcsystem}. This Figure illustrates different boundaries, regions and structures that populate them. They correspond to the observed phenomena shown in Figure~\ref{fig:sfc}a. The magnetic field line that barely touches the bow-shock is called the tangential IMF line. Just downstream of it the spacecraft would first detect reflected electrons, so this region is called the electron foreshock (green). Further downstream, where magnetic field lines connect to the quasi-parallel bow-shock, begins the ion foreshock (yellow). There are no ULF waves in this region and the reflected ions follow IMF lines, so they are called field-aligned ion beams (FAB). Still further downstream transverse ULF waves are also observed and this is where the ULF wave foreshock begins (purple). This region is delimited by a thick dash-dotted line that corresponds to the ULF wave boundary \citep{greenstadtbaum86} also shown in Figure~\ref{fig:sfc}a. In this region observed ion distributions change from FAB to intermediate. The thick dashed line corresponds to intermediate ion boundary \citep[see Figure~\ref{fig:sfc}a and][]{mezianeduston98}. Finally, a spacecraft crosses an FCB (thick dotted line) and enters the region (blue) with compressive ULF waves, shocklets, SLAMS, diffuse suprathermal ions and other transient structures such as foreshock cavitons and SHFAs.

Another way to compare different phenomena is to look at the IMF and SW conditions under which they are observed. Figure~\ref{fig:swdist}a shows distributions of (from left to right) IMF magnitude, SW density, SW velocity and SW thermal pressure for (from top to bottom) SHFAs, foreshock cavitons, foreshock cavities and FCBs. Figure~\ref{fig:swdist}b is in the same format but it shows distributions of (from left to right) SW temperature, IMF cone angle $\theta_{BX}$, SW Alfv\'en velocity V$_A$ and Alfv\'enic Mach number M$_A$. No SHFAs were observed for M$_A<$6 and V$_A>$90~kms$^{-1}$, although a larger sample should be analyzed to reach any definite conclusions. Distributions of SW and IMF properties in the case of cavities, cavitons and FCBs are much more similar, although the ranges of B magnitud, plasma densities tend to be larger for cavities and cavitons.
Although the distributions presented in this Figure are subject to the intrinsic distributions of the IMF and SW properties, we can see that all four types of upstream transient structures may be observed under a wide range of SW and IMF conditions.

\section{Discussion}

In the first part of this paper we use multi-spacecraft data from the Cluster and THEMIS missions to confirm some predictions from \citet{omidi13a} hybrid simulations, namely the existence of a traveling foreshock and the relation between some of the phenomena that are commonly observed upstream of the Earth's bow-shock.

We postulate here that two types of foreshock may exist upstream of the Earth's bow-shock: one is a global Earth's foreshock that forms upstream of a quasi-parallel section of the Earth's bow-shock. It was shown by \citet{omidi13a} that during steady solar wind and IMF conditions a foreshock compressional boundary forms at the edge of a foreshock, delimiting a region of either pristine solar wind or a region of field-aligned ion beams from a region of diffuse ions that is populated by compressive ULF waves. In practice the SW and the IMF are never exactly steady. IMF rotations are commonly observed in the solar wind \citep[e.g., ][]{borovsky08}. Such rotations may be slow, lasting for several minutes (as in our case study 1), or they can occur on very short times ($\sim$seconds). The latter are called rotational discontinuities. When two consecutive IMF rotations pass the Earth's bow-shock, the foreshock changes its position with respect to the bow-shock and may undergo back and forth motion. This can cause a spacecraft to enter the foreshock for a time period that can range from a few minutes to some tens of minutes and then exit it.

Foreshock observations on similar time scales can also occur when bundles of magnetic field lines with orientations different from the rest of the IMF sweep along the bow-shock surface. In these cases a spacecraft located upstream of the shock observes two consecutive IMF RDs. The rotation of the IMF across such two RDs may be sufficient to temporarily change the geometry of a portion of the bow-shock from quasi-perpendicular to quasi-parallel. The region of space between the two RDs becomes populated by suprathermal ions and compressive ULF fluctuations and resembles a global foreshock. However, because the RDs are convected by the SW antisunwards, their intersection with the bow-shock propagates along the bow shock surface in the direction roughly perpendicular to Sun-Earth line. The foreshock-like region between the RDs then also propagates in the same direction hence we call it a traveling foreshock.

Foreshock compressional boundaries form at the edges of both types of foreshock when they are observed under unsteady IMF conditions. FCBs are observed in the B and N profiles simultaneously with either slow rotations of the IMF or with IMF RDs. IMF rotation across FCBs has been analyzed by \citet{RojasCastillo13}.  In their statistical study of FCBs these authors 
reported that the IMF cone angle changed by up to 15$^\circ$ across 36~\% of their events, while it changed between 15$^\circ$ and 30$^\circ$ across 42~\% of the events. However \citet{RojasCastillo13} did not look at whether or not all their FCBs were related to IMF slow rotations and IMF RDs. According to numerical simulations of \citet{omidi13a} B-field rotations across FCBs occur even under steady IMF.

We further point out that other transient structures, isolated foreshock cavities, also form due to two successive IMF RDs. However in the case of isolated cavities the two RDs are very closely separated, so the ULF waves are either not observed in the region between them or they are few. We conclude that these cavities are a subset of traveling foreshocks. The appearance of traveling foreshocks varies as the separation between the bounding IMF RDs increases. As the separation increases, the ULF waves begin to appear in the region between them. We illustrate this by showing a case study where only about ten ULF waveforms are observed during the time between two IMF RDs.

The rotations of IMF can be easily understood if one imagines the IMF to be composed of magnetic flux tubes that extend from the solar surface into the interplanetary space. \citet{borovsky08} performed a statistical study of flux tubes properties. These authors studied IMF rotations at 1~AU from 1998 to 2004 and concluded that small rotations with characteristic rotation angles of 15$^\circ$ occur due to IMF turbulence, while larger rotations occur when different magnetic flux tubes are convected across the observer. \citet{borovsky08} estimated that the thicknesses of these flux tubes at heliocentric distance of 1~AU range from less than 10~R$_E$ to several thousands of R$_E$ with median sizes being 98~R$_E$ and 67~R$_E$ for the slow and fast solar wind, respectively. These flux tubes exhibit very different durations in the spacecraft data. It is also suggested in the sketch in Figure~1 of \citet{borovsky08} that the flux tubes are neither straight nor are they simply aligned along the Parker spiral, but that they change their orientation in space. They can be distorted with wiggles, and they are interlaced. It is thus appropriate to suggest that slow IMF rotations, such as the one observed in our case study 1 (the global foreshock) occur when a large 
flux tube with a wiggle is being convected pass the observer. We illustrate this idea in Figure~\ref{fig:sketch20070807}c where a large magnetic flux tube is colored in red. The sketched flux tube exhibits a kink and is convected antisunwards (downwards in the Figure) by the solar wind. As the kink passes the bow-shock the latter remains inside this flux tube but the orientation of the B-feld changes with time. In this case the spacecraft detect back and forth motion of the foreshock due to varying angle $\theta_{BN}$.

Fast rotations and RDs (related to traveling foreshocks) on the other hand are observed as different flux tubes convect across the observer. We show such a situation in Figure~\ref{fig:sketch20070814}b where the bow-shock is initially inside the large kinked flux tube (red). At some point either the whole bow-shock or just a portion of its surface, briefly exits the red flux tube and enters a thinner flux one (blue) with different IMF orientation. In this case, multiple spacecraft detect the convecting foreshock signature.

We should note here that in case of very parallel IMF, the related RDs would also appear slow rotations since it would take a long time for a spacecraft to cross from one flux tube to another.

We also look at phenomena inside the traveling foreshock. We observe compressive ULF waves and in one case a spontaneous hot flow anomaly. This means that other structures may also form inside such foreshocks: compressive ULF waves may evolve into short-large amplitude magnetic structures \citep[SLAMS, e.g.][]{thomsen90, schwartzburgess91, schwartz92, mann94, lucek02} which may cause cyclic reformation of a portion of the Earth's bow-shock. The interaction of compressive and transverse ULF waves leads to the formation of foreshock cavitons \citep{omidi07} and interactions of cavitons with the bow-shock lead to formation of the SHFAs \citep[][]{zhang13, omidi13b} and further to the rippling of the bow-shock's surface. On the other hand, hybrid simulations suggest that foreshock cavitons and SHFAs may temporarily and locally weaken the bow-shock, so its transition exhibits smaller B-field magnitudes and densities (Blanco-Cano et al., in preparation).

Rippling and weakening of the bow-shock may have consequences in the magnetosheath. Specifically, these processes have been identified as formation mechanisms for magnetosheath jets \citep[e.g.][]{hietala09}.
Jets have mostly been detected downstream of the quasi-parallel bow-shock. IMF RDs are associated with traveling foreshocks so the magnetosheath jets could sometimes appear in association with them.

In the second part of this paper we statistically compare observational properties of four foreshock phenomena (foreshock cavities, foreshock cavitons, foreshock compressional boundaries and spontaneous hot fow anomalies), their observed locations and the SW and IMF conditions under which they were detected. All of these phenomena show changes in the B-field magnitude and plasma density when compared to the conditions of the ambient medium. All but FCBs show depletions of these two quantities in their centers. We show that SHFAs are the most depleted structures inside which the B-field and N diminish typically by $\sim$90~\%. In the case of cavities and cavitons this number is between 40~\% and 50~\% on average. The changes in magnetic field and in density are most correlated in the case of foreshock cavitons and FCBs with correlation coefficients of 0.85 and 0.86, respectively.
Strong depletions in the case of SHFAs are expected following the proposed explanation for their formation, namely that they occur due to interactions of already depleted structures in the foreshock, the foreshock cavitons, with the bow-shock. When this interaction occurs, the ions at their centers energize due to ion trapping by the cavitons and ion reflection between the bow-shock and the cavitons and this leads to further depletion of B and N inside them \citep{zhang13, omidi13b}.

By comparing locations of the four phenomena in solar foreshock coordinates we show that FCBs and foreshock cavities (or traveling foreshocks) occupy the same domain, which strengthens the proposal first made by \citet{omidi13a} that the two phenomena are related, namely that FCBs occur at the cavities's edges. It should be pointed out here that the FCBs that occur at the edges of the global foreshock would also occupy the same domain. In SFC the FCBs and foreshock cavities are the phenomena that appear upstream of ULF waves, intermediate ion boundaries (related to global foreshock) while foreshock cavitons appear downstream of them. This makes sense since foreshock cavitons are the result of the interaction of transverse and compressive waves ULF waves and the compressive ULF wave appear further inside the foreshock than the ULF wave boundary studied by \citet{greenstadtbaum86}. On the other hand the FCBs and traveling foreshocks are bounded by pristine solar wind which will positioned them upstream of all the other phenomena in the SFC coordinates.

Finally we show that all four phenomena occur for a wide range of SW and IMF conditions.

\section{Conclusions}
Here we summarize the conclusions of this investigation:
\begin{itemize}
\item There are two different types of foreshock detected upstream of the Earth's bow shock. One is the global foreshock located upstream of quasi-parallel section of the bow-shock. This foreshock may change its location due to IMF rotations. Two succesive IMF rotations may cause the back and forth motion of the foreshock resulting in brief excursions of the spacecraft into it that can last between several tens of minutes to several hours. Another type is the traveling foreshock which exists between two IMF RDs. This kind of foreshock usually lasts of the order of ten minutes in the spacecraft data. We call it a traveling foreshock since it propagates along the bow-shock surface. We should stress out though that when the flux tube is large enough, it can affect the whole bow shock surface so the resulting traveling foreshock will also be ``global''.
\item The difference between what is traditionally called the global foreshock and the traveling foreshock is not their size. In the case of back and forth motion of the global foreshock, the orientation of the IMF changes, but the bow-shock remains inside the same flux tube. In case of traveling foreshock the bow-shock (or a portion of it) magnetically connects to different flux tubes.
\item Foreshock compressional boundaries are observed at the edges of either type of foreshock. 
\item Foreshock cavities are a subset of traveling foreshocks, where the two IMF RDs are so close that ULF waves are either not observed or only few of them are observed between the RDs. All the isolated foreshock cavities in the literature exhibit durations of less than 200 seconds in the data, while the traveling foreshocks can last for ten or more minutes. We must however permit the possibility that on rare occasions the signatures in the spacecraft data very similar to those of foreshock cavities could also be observed due to the spacecraft brief encounters with the global foreshock. 
\item Compressive ULF waves and transient foreshock structures inside traveling foreshocks can cause bow-shock reformation and rippling. In the past shock rippling has been proposed as a formation mechanism for magnetoshetath jets. These should then also appear in association with traveling foreshocks.
\item Foreshock transient structures, such as spontaneous hot flow anomalies have been shown to exist inside the traveling foreshocks. According to present knowledge, the SHFAs are a product of foreshock cavitons interacting with the Earth's bow-shock. Hence, it should in principle be possible to observe foreshock cavitons inside traveling foreshocks.
\item SHFAs are the most depleted structures in terms of B-field and plasma density inside their cores when compared to the surrounding medium.
\item The changes in plasma density and B-field are most correlated in case of FCBs and foreshock cavitons.
\item The FCBs and foreshock cavities occupy the same domain in SFCs, which agrees with the idea that the FCBs form at the edges of the cavities (or traveling foreshocks).
\item Foreshock cavities, cavitons, FCBs and SHFAs can be observed under a wide range of SW and IMF conditions.
\end{itemize}

Some challenges remain for future work. Foreshock cavitons, have not yet been observed inside traveling foreshocks. We only find SHFAs inside these foreshocks and infer that cavitons must also exist there. Similarly, we found compressive ULF waves but did not  look for shocklets and SLAMS inside traveling foreshocks. Finally, simultaneous observations of traveling foreshock and magnetosheath jets would provide a conclusive piece of evidence of their possible relation.

\acknowledgments
The authors are grateful to Dr. S. J. Schwartz for providing the processed observational data for foreshock cavities and to the Cluster Science Archive teams and the CL/CLWeb team for the easy access to the Cluster and THEMIS data. The data used in this work can be accessed on the corresponding webpages: http://clweb.irap.omp.eu and https://www.cosmos.esa.int/web/csa. This work has been supported by the International Space Science Institute (ISSI). PK's work was also supported by the PAPIIT grant IA104416. X.B.C. and P.K. work was supported by DGAPA/PAPIIT grant IN105014 and CONACYT grant 179588. Work at GSFC was supported by the THEMIS mission. 
 
\email{aastex-help@aas.org}.

{\it Facilities:} \facility{CSA}.

\pagebreak

\clearpage

\begin{table}
\begin{tabular}{ccc}
\hline
\hline
Date& Start time [UT] & End Time [UT]\\
yyy-mm-dd & hh:mm:ss & hh:mm:ss\\
\hline
2003-03-10 & 21:28:46 & 21:29:09 \\
2003-04-12 & 00:22:56 & 00:23:52 \\
2003-04-27 & 17:44:48 & 17:45:20 \\
2003-04-30 & 03:41:35 & 03:42:27 \\
2003-04-30 & 19:38:27 & 19:39:17 \\
2006-03-10 & 15:30:26 & 15:31:05 \\
2006-03-21 & 08:48:52 & 08:49:13 \\
2006-05-23 & 01:58:49 & 01:58:58 \\
2008-03-14 & 11:43:25 & 11:45:32 \\
2008-04-07 & 02:57:22 & 02:58:15 \\
2008-04-08 & 20:27:39 & 20:28:23 \\
2009-03-19 & 14:37:09 & 14:38:20 \\
2009-03-19 & 14:39:36 & 14:41:25 \\
2009-03-30 & 12:39:32 & 12:41:44 \\
2009-04-07 & 14:16:15 & 14:16:25 \\
2009-04-23 & 04:47:38 & 04:48:40 \\
2009-05-08 & 14:23:19 & 14:23:51 \\
2011-02-07 & 10:13:10 & 10:14:20 \\
2011-03-05 & 11:36:01 & 11:38:01\\
\hline 
\end{tabular}
\caption{List of Spontaneous Hot Flow Anomalies.}
\label{tab1}
\end{table}

\begin{figure*}
\includegraphics[width = 0.9\textwidth]{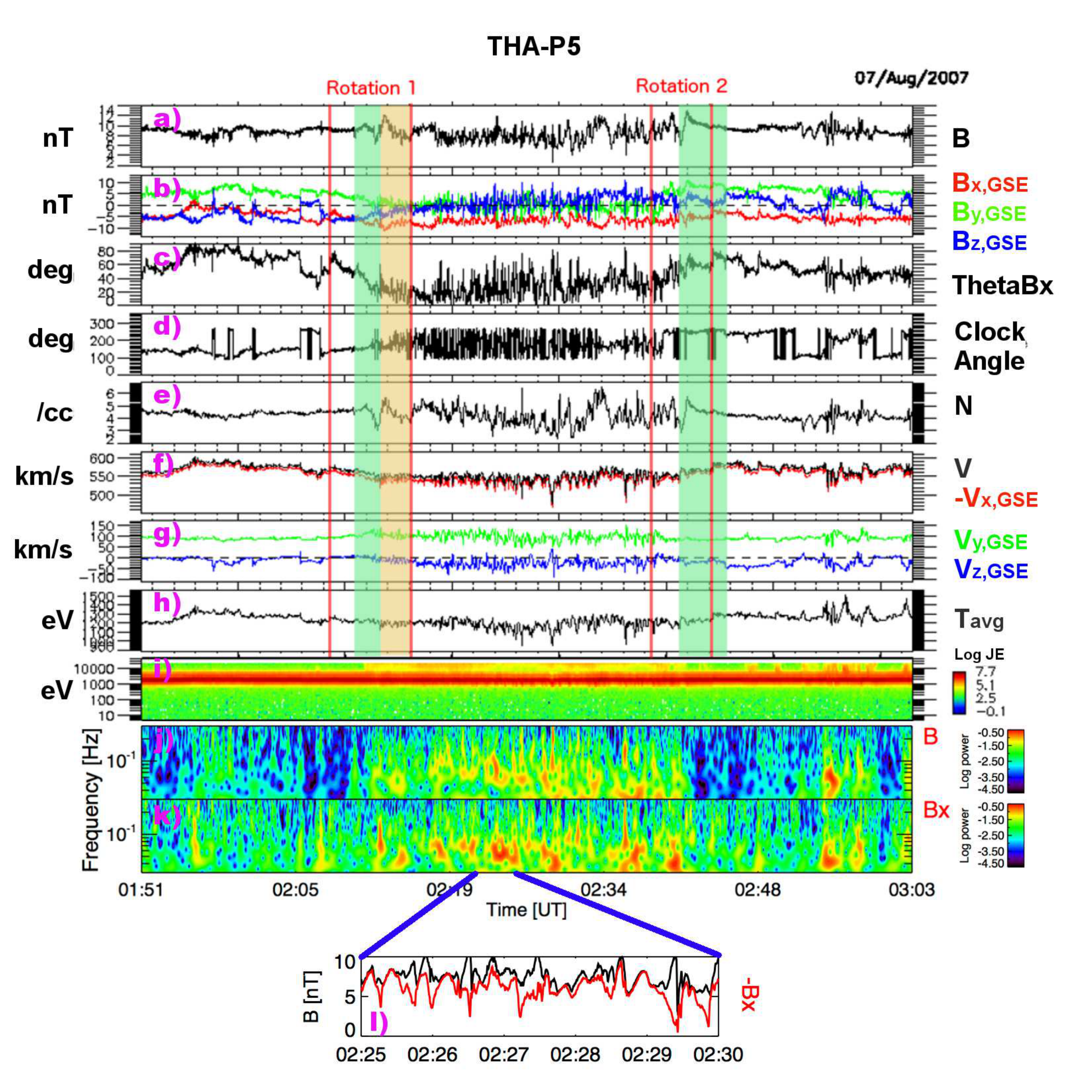}
\caption{THEMIS A data between 01:51~UT and 03:03~UT on 7 August 2007. Panels show: a) magnetic field magnitude in units of nanoTesla (nT), b) magnetic field components in GSE coordinate system in units of nT, c) angle between the IMF and the Sun-Earth line in degrees, d) IMF clock angle in degrees, e) SW density in cm$^{-3}$, f) solar wind speed (black) and -V$_x$ component (red) in kms$^{-1}$, g) V$_y$ and V$_z$ components of SW velocity in kms$^{-1}$, h) SW temperature in eV, i) ion spectra with colors representing the logarithm of the particle energy flux (units eV/(cm$^-1\cdot$s$\cdot$str$\cdot$eV)), (j) Morlet wavelet spectrum for B-magnitude and (k) Bx-component and l) B-magnitude and Bx component between 02:25-02:30~UT.}
\label{fig:20070807a}
\end{figure*}

\begin{figure*}
\includegraphics[width = 0.8\textwidth]{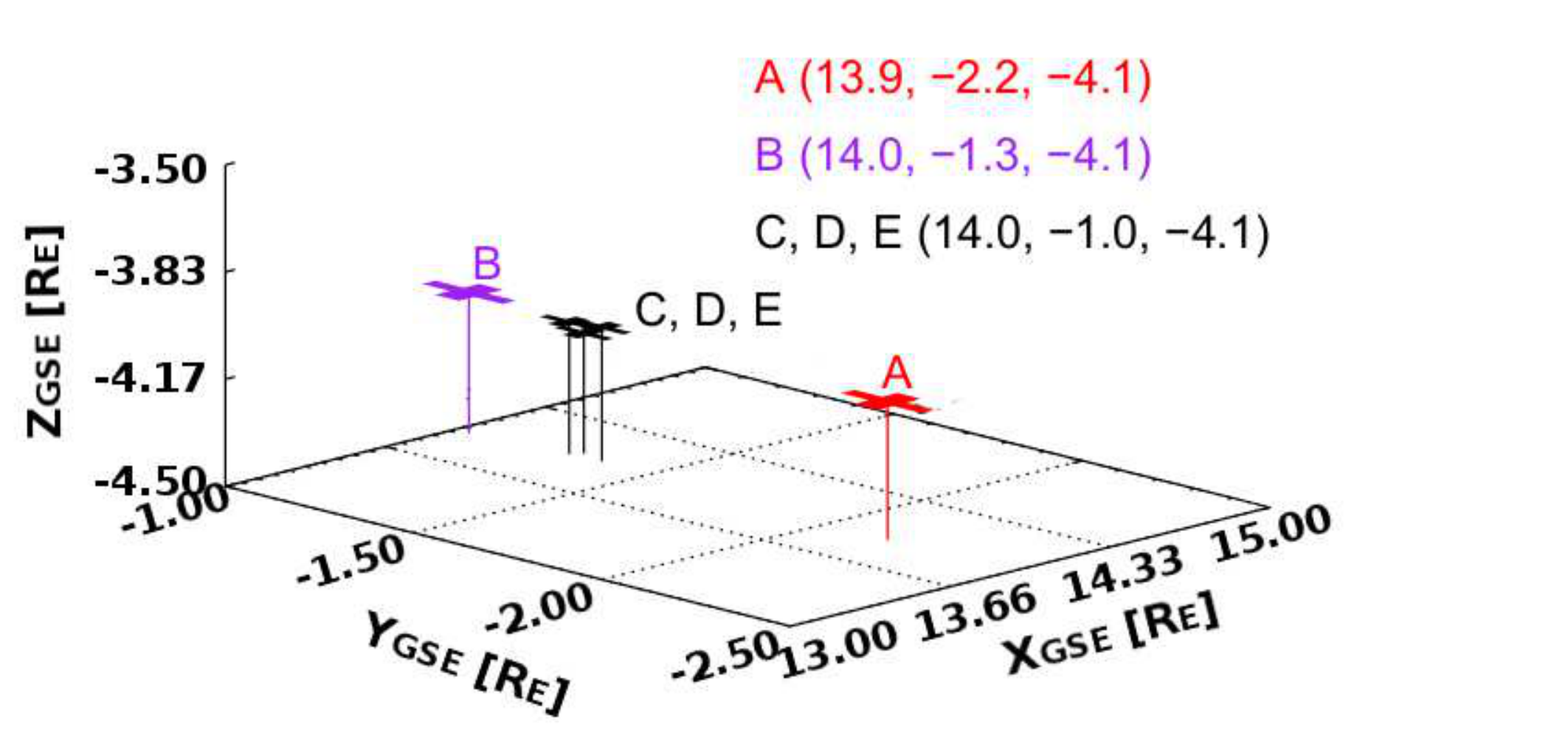}
\caption{Spatial configuration of the five THEMIS spacecraft during the 7 August 2007 event.}
\label{fig:20070807b}
\end{figure*}

\begin{figure*}
\centering
\includegraphics[width = 0.6\textwidth]{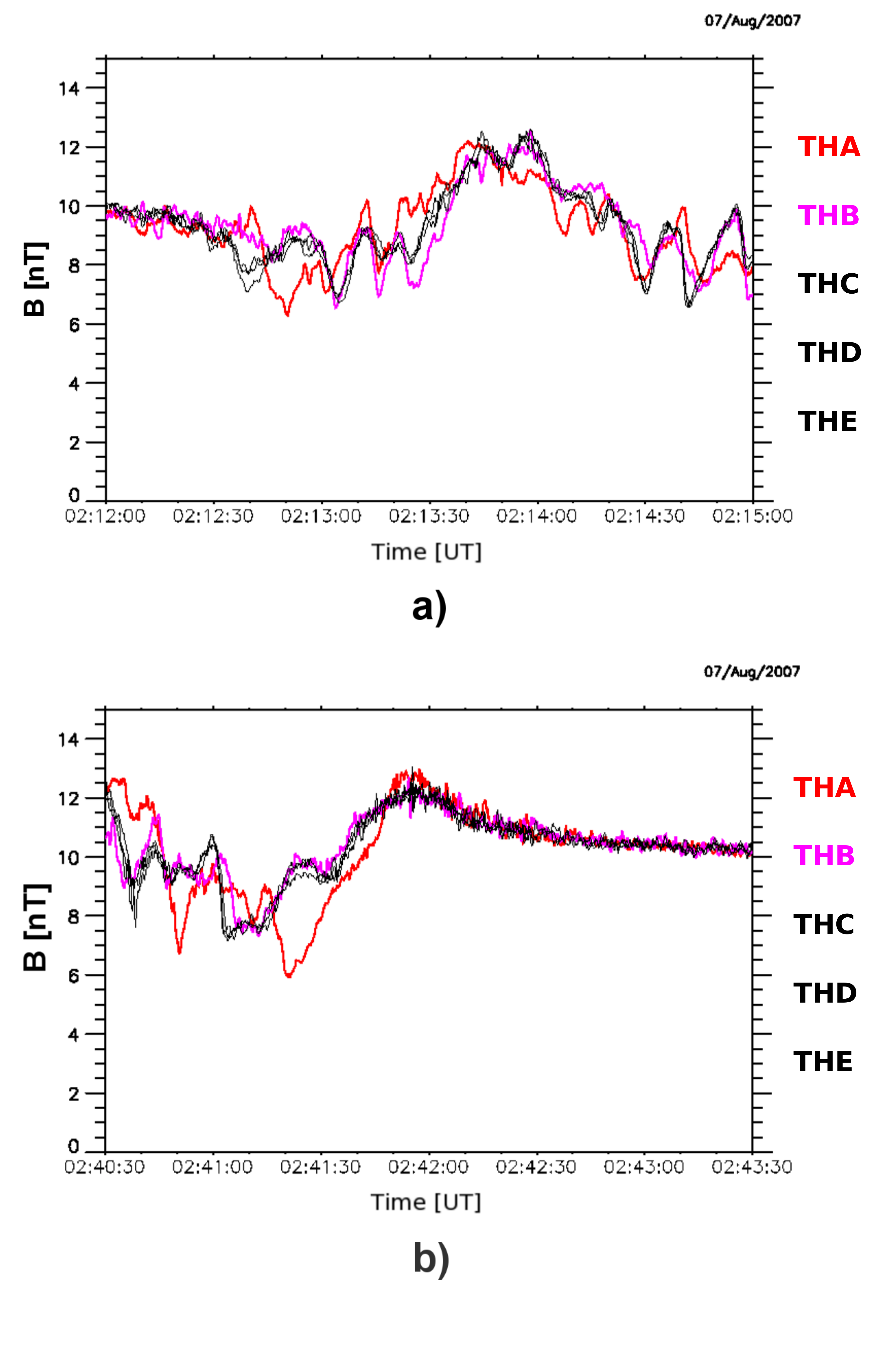}
\caption{Magnetic field magnitude profiles of the leading (a) and the trailing edge (b) associated to the foreshock detected on 7 August 2007. Red line represents the THEMIS A data, purple line the THEMIS B data, while the data of the other three spacecraft are represented by the black traces.}
\label{fig:20070807c}
\end{figure*}

\begin{figure*}
\centering
\includegraphics[width = 0.90\textwidth]{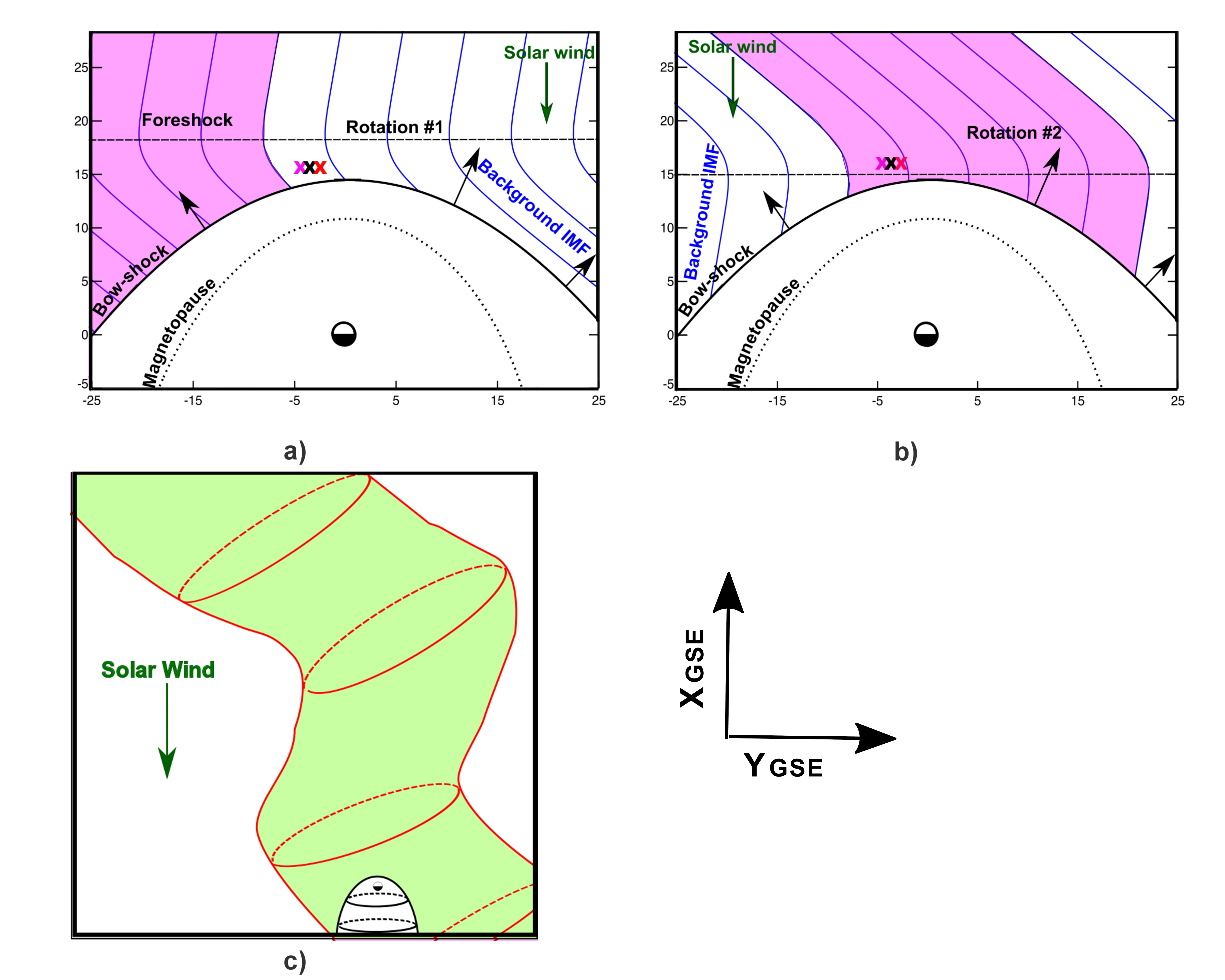}
\caption{(a) and (b) Sketch of interplanetary magnetic field and foreshock configurations just before the interaction of two IMF rotations with the Earth's bow-shock. \citet[][]{fairfield71} models for bow-shock and magnetopause have been used here. The black arrows represent the local directions of the bow-shock normal. The crosses represent spacecraft in a configuration similar to that in Figure~\ref{fig:20070807b}. c) A large wiggled magnetic flux tube passing by the bow shock would cause the observer to detect slowly rotating IMF and the nonconvecting (back and forth) motion of the foreshock as the $\theta_{BN}$ at every point on the bow-shock surface changes with time. It should be stressed out that what is shown is just one scenario since in reality the properties of the flux tube, such as its extension, width and spatial location can be very different from the ones shown here.}
\label{fig:sketch20070807}
\end{figure*}

\begin{figure*}
\includegraphics[width = 0.9\textwidth]{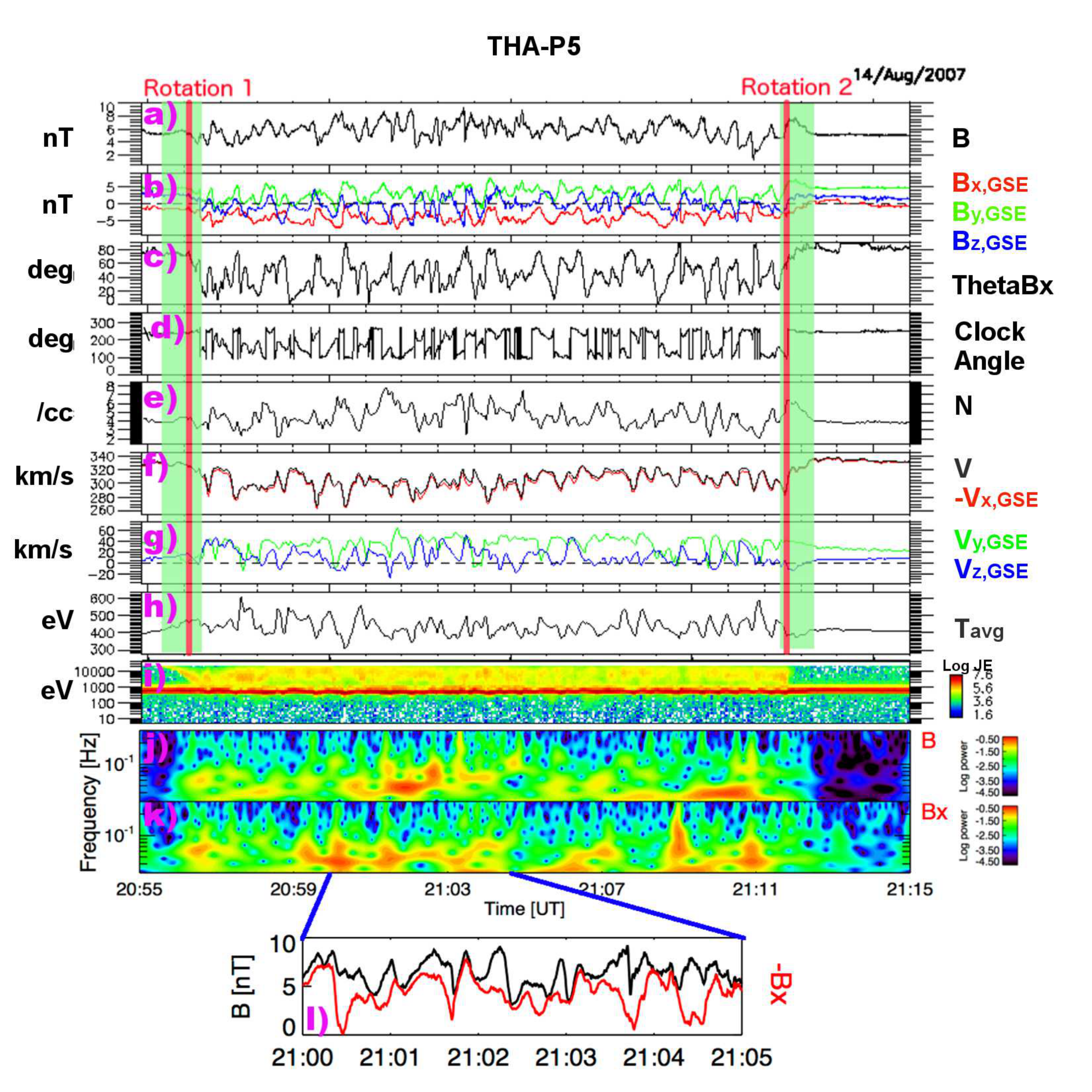}
\caption{a) THEMIS A data between 20:55~UT and 21:16~UT on 14 August 2007. The figure is in the same format as Figure~\ref{fig:20070807a}. The two vertical red lines show two IMF RDs and the intervals shadowed in green mark the FCBs at the edges of the traveling foreshock.}
\label{fig:20070814a}
\end{figure*}

\begin{figure*}
\includegraphics[width = 0.8\textwidth]{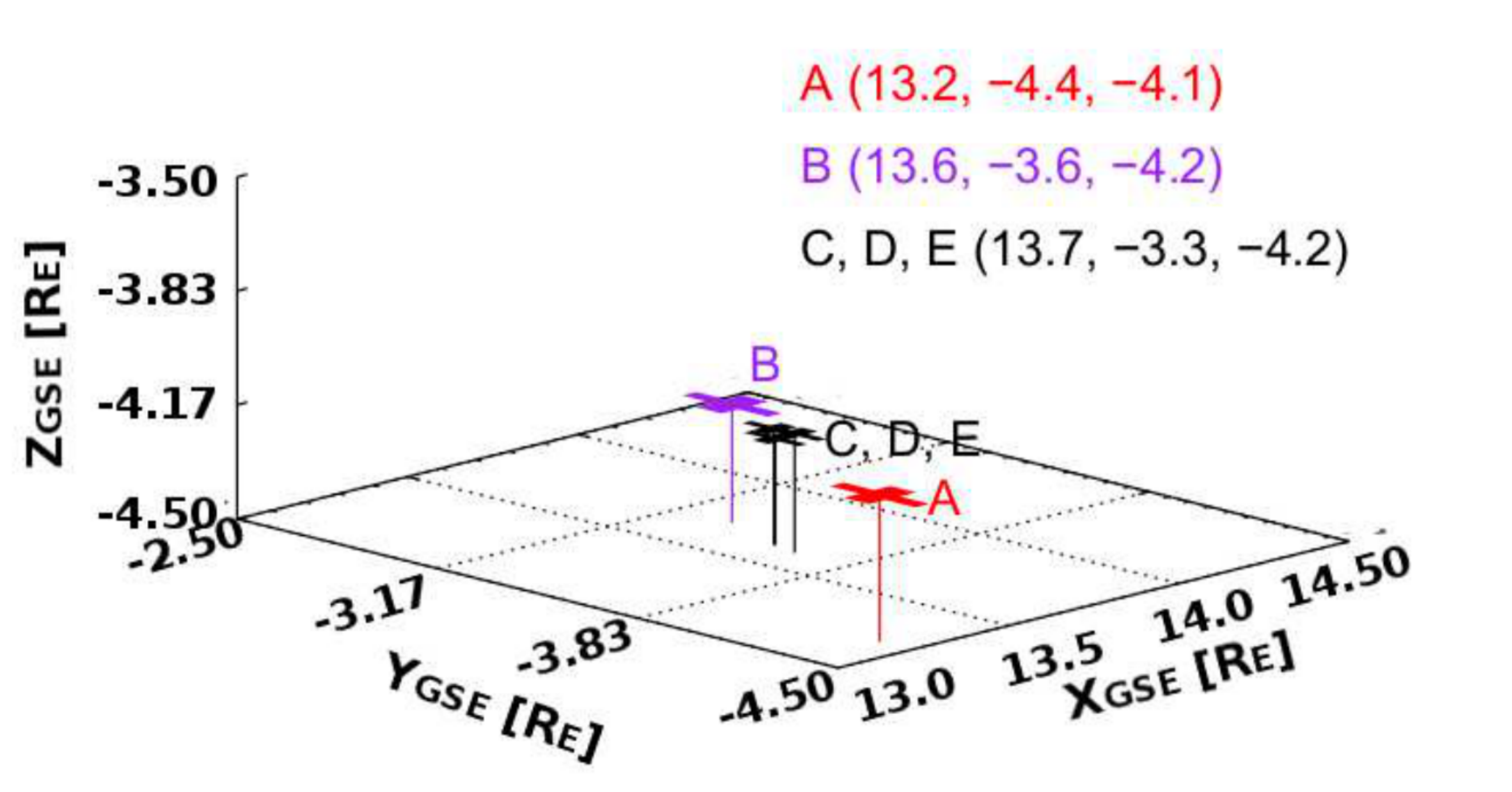}
\caption{Spatial configuration of the five THEMIS spacecraft during the 14 August 2007 observations of the traveling foreshock.}
\label{fig:20070814b}
\end{figure*}

\begin{figure*}
\centering
\includegraphics[width = 0.5\textwidth]{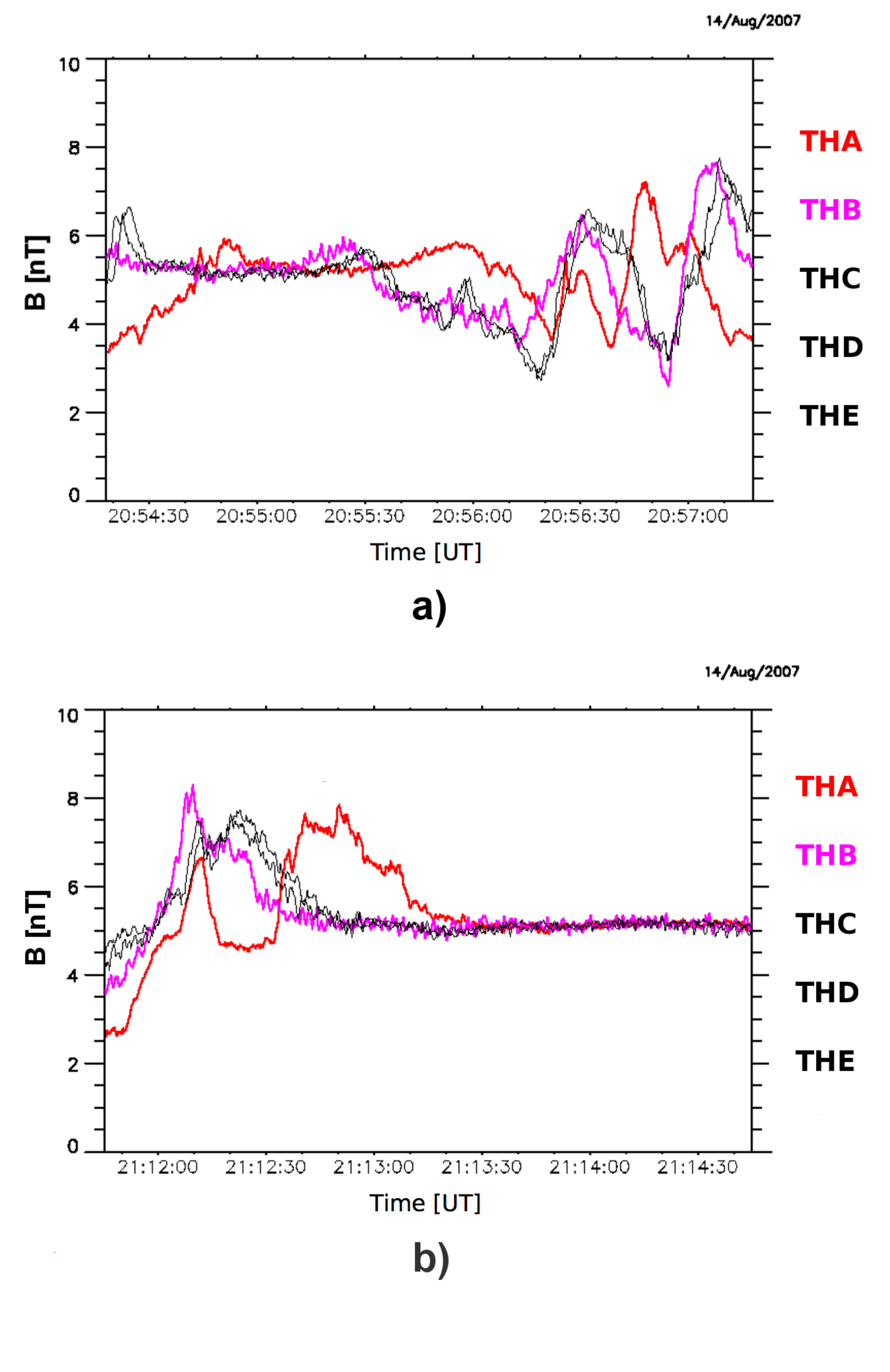}
\caption{Magnetic field magnitude profiles of the leading (a) and trailing (b) edges of the foreshock detected on 14 August 2007. Red trace represents the THEMIS A data, purple trace the THEMIS B data, while the data of the other three spacecraft are represented by the black traces. Vertical lines mark times of the FCB peak, stated in the text.}
\label{fig:20070814c}
\end{figure*}

\begin{figure*}
\centering
\includegraphics[width = 0.40\textwidth]{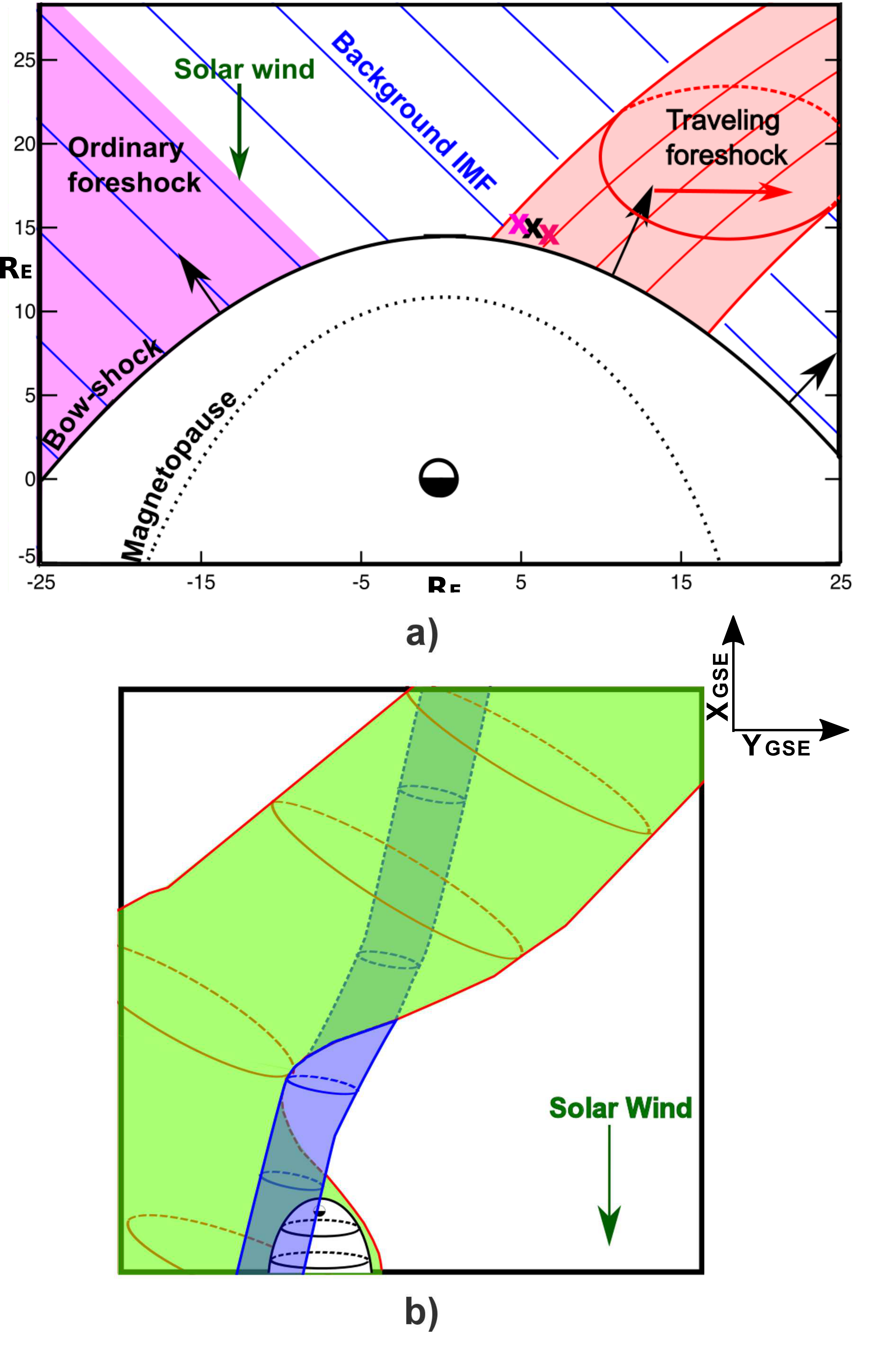}
\caption{a) Sketch of the event observed on 14 August 2007 (\citet[][]{fairfield71} models for bow-shock and magnetopause have been used here). The purple color represents the global foreshock and the blue lines represent the IMF. The flux tube with different orientation than the background IMF is colored in red. The black arrows are the local bow-shock normals. The red arrow determines the propagation of the traveling foreshock, while the green arrow shows the SW propagation direction. The crosses represent spacecraft in a configuration similar to that in Figure~\ref{fig:20070814b}. b) Due to twisted and braided magnetic field lines the Earth's bow-shock passes through different magnetic flux tubes. Here the bow-shock is first inside the red flux tube, then it passes through a blue tube and returns into the red one. This results in rapid changes in the IMF orientation. As spacecraft cross an interface between two flux tubes they observe a rotational discontinuity. Due to different IMF orientations inside the flux tube, spacecraft observe the convected foreshock signature.}
\label{fig:sketch20070814}
\end{figure*}

\begin{figure*}
\centering
\includegraphics[width = 0.7\textwidth]{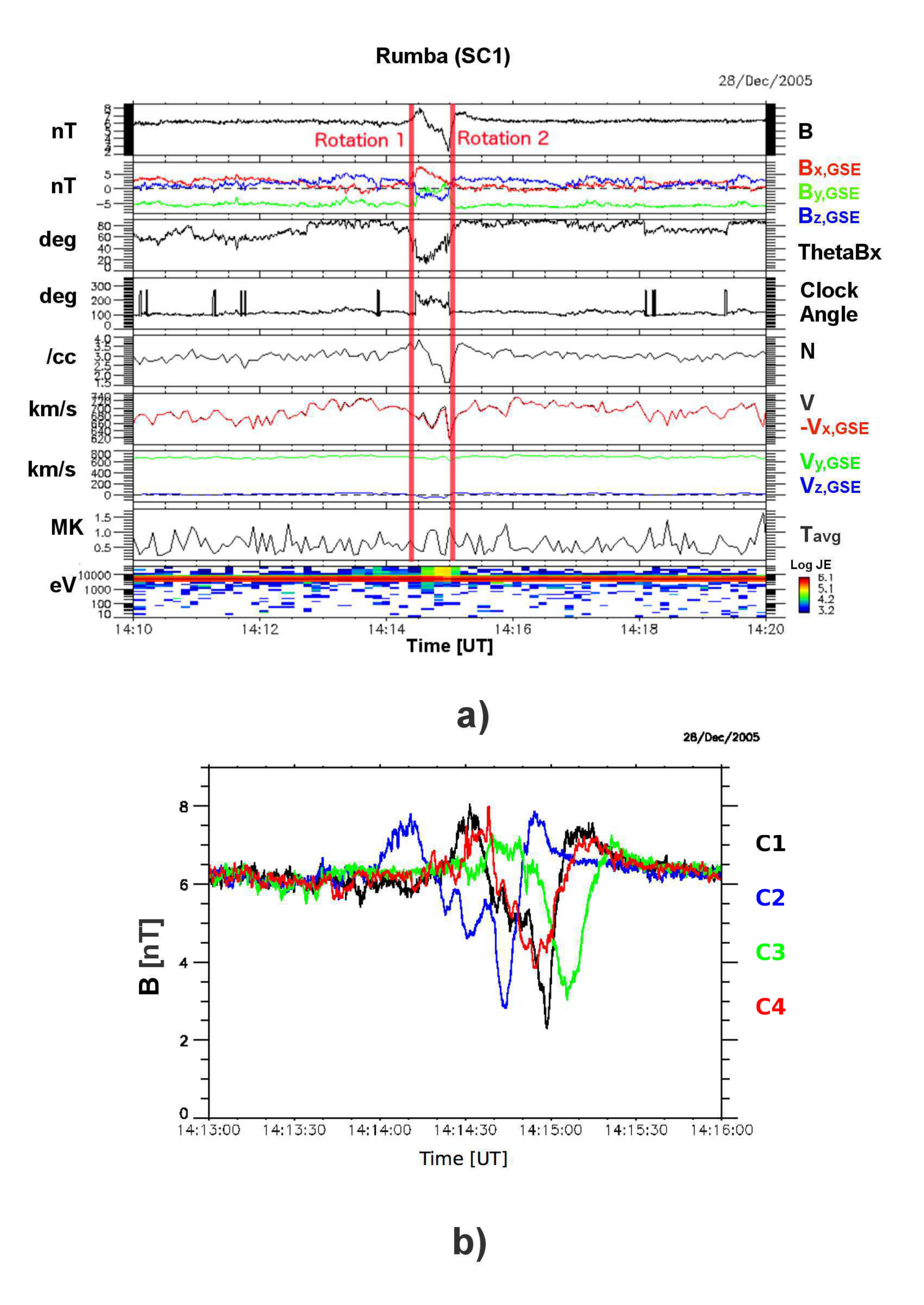}
\caption{Foreshock cavity. a) Cluster~1 data showing between 14:10~UT and 14:20~UT on 28 December 2005. The Figure is in the same format as Figure~\ref{fig:20070807a} except that the temperature is in units of megaKelvins (MK) and we do not show any wavelet spectra. b) Magnetic field magnitude profiles of the four Cluster spacecraft during the 28 December 2005 event. The C1, C2, C3 and C4 data are represented by the black, blue, green and red traces, respectively.}
\label{fig:20051228a}
\end{figure*}

\begin{figure*}
\centering
\includegraphics[width = 0.7\textwidth]{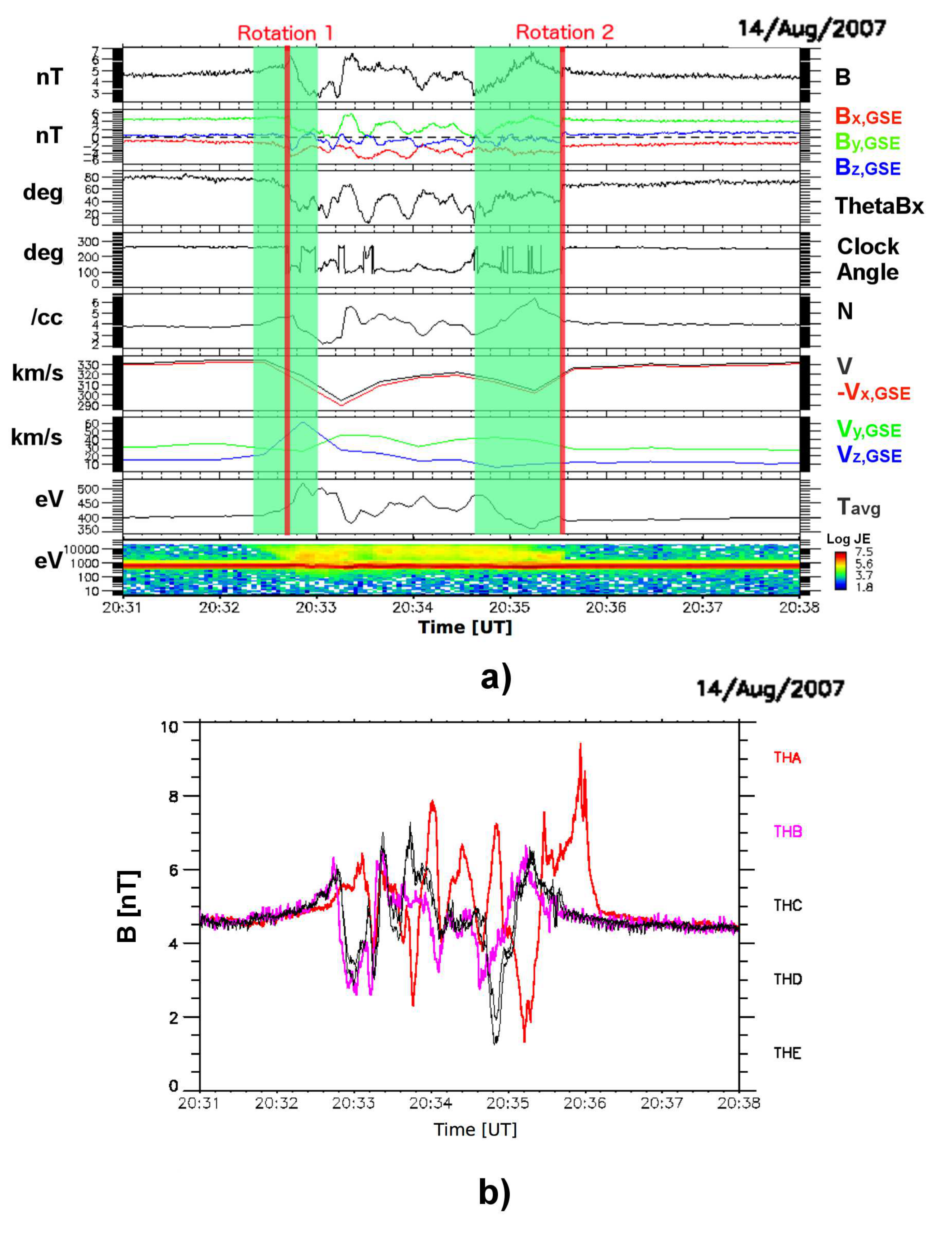}
\caption{Foreshock cavity. a) THEMIS~A data between 20:31~UT and 20:38~UT on 14 August 2007. The Figure is in the same format as Figure~\ref{fig:20051228a}a). The vertical red lines mark the IMF RDs and the intervals shadowed in green mark the FCBs. b) Magnetic field magnitude profiles of the five THEMIS spacecraft during the 14 august 2007 intermediate event. Red line represents the THEMIS A data, purple lie the THEMIS B data, while the data of the other three spacecraft are represented by the black traces.}
\label{fig:20051228c}
\end{figure*}

\begin{figure*}
\centering
\includegraphics[width = 1.0\textwidth]{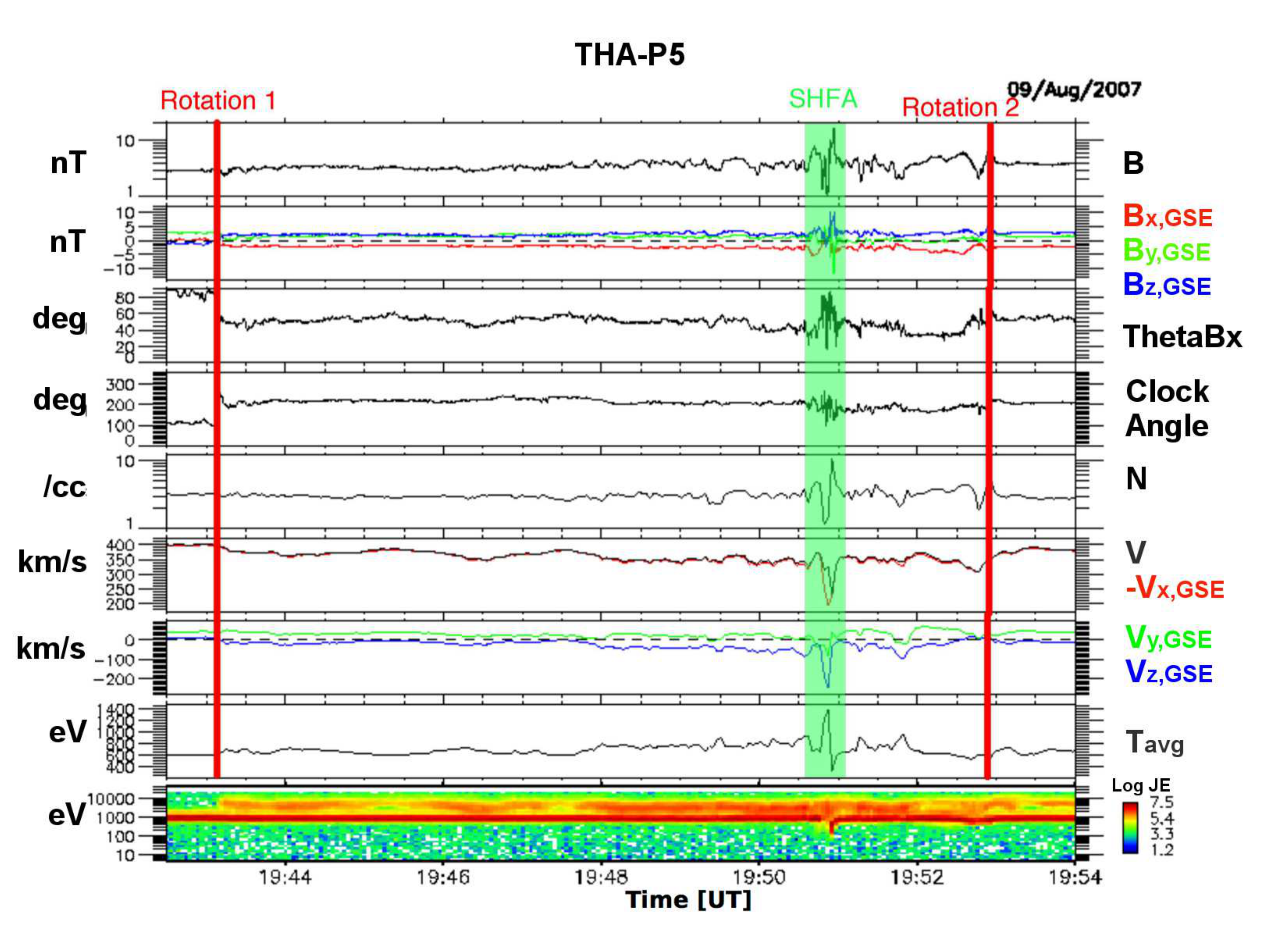}
\caption{THEMIS A data between 19:42:30~UT and 19:54~UT on 9 August 2007. The figure is in the same format as Figure~\ref{fig:20051228a}a) except that B magnitude and plasma density N are represented on a logarithmic scale in order to bring out the compressive ULF waves in the traveling foreshock. The two vertical red lines show two IMF RDs delimiting the traveling foreshock, while the intervals shadowed in green shows the spontaneous hot flow anomaly.}
\label{fig:20070809}
\end{figure*}

\begin{figure*}
\centering
\includegraphics[width = 1.\textwidth]{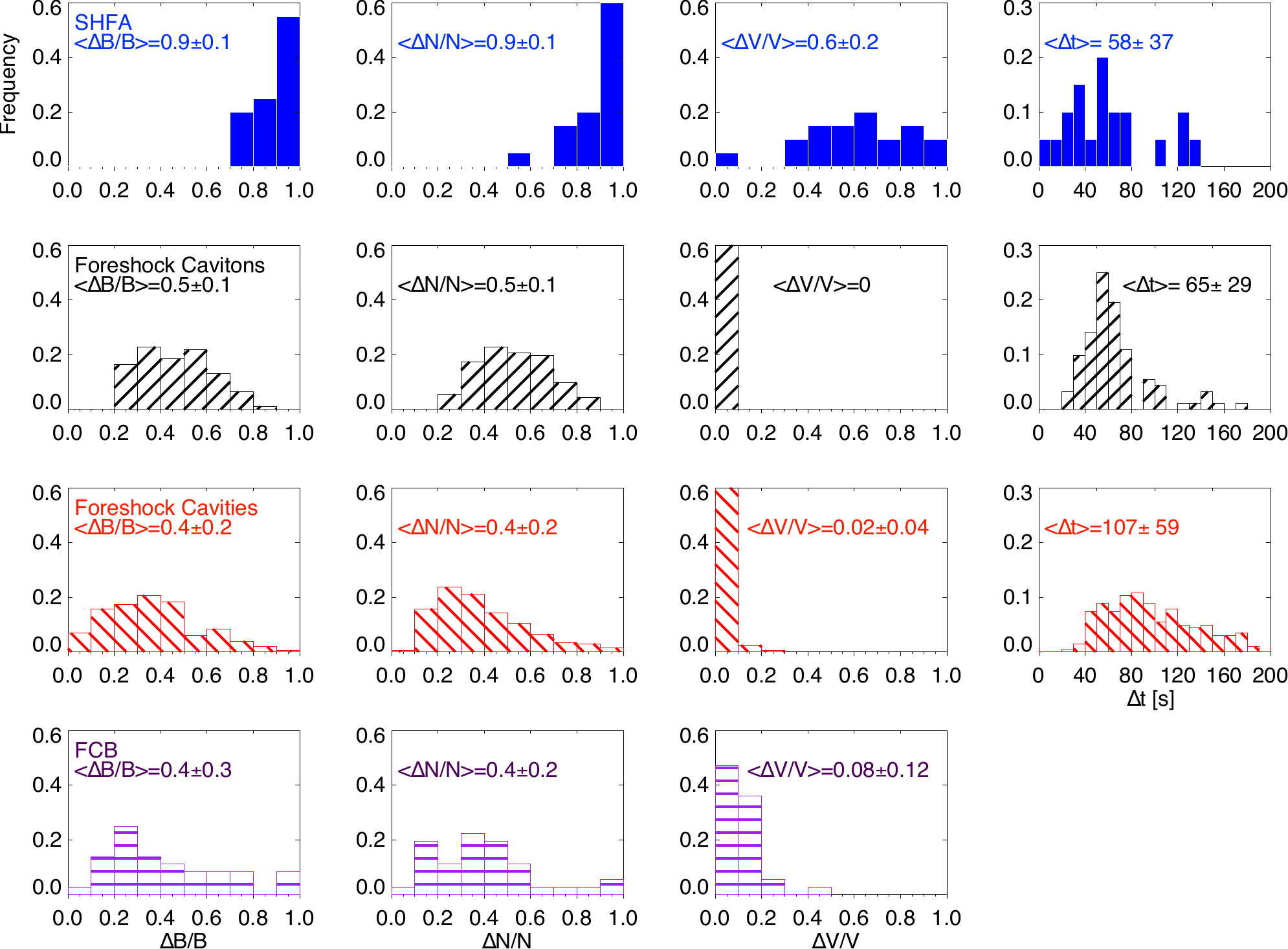}
\caption{Some statistical distributions of observational properties of (from top to bottom) SHFAs, foreshock cavitons, foreshock cavities and FCBs. The following quantities are shown: relative changes of (from left to right) magnetic field magnitude, density and plasma velocity and durations. The $\Delta$ sign marks the difference between the ambient SW value and the minimum value inside the structures. In case of FCBs it represents the difference between the maximum value inside the FCB and the upstream SW value.}
\label{fig:distributions}
\end{figure*}

\begin{figure*}
\centering
\includegraphics[width = 0.5\textwidth]{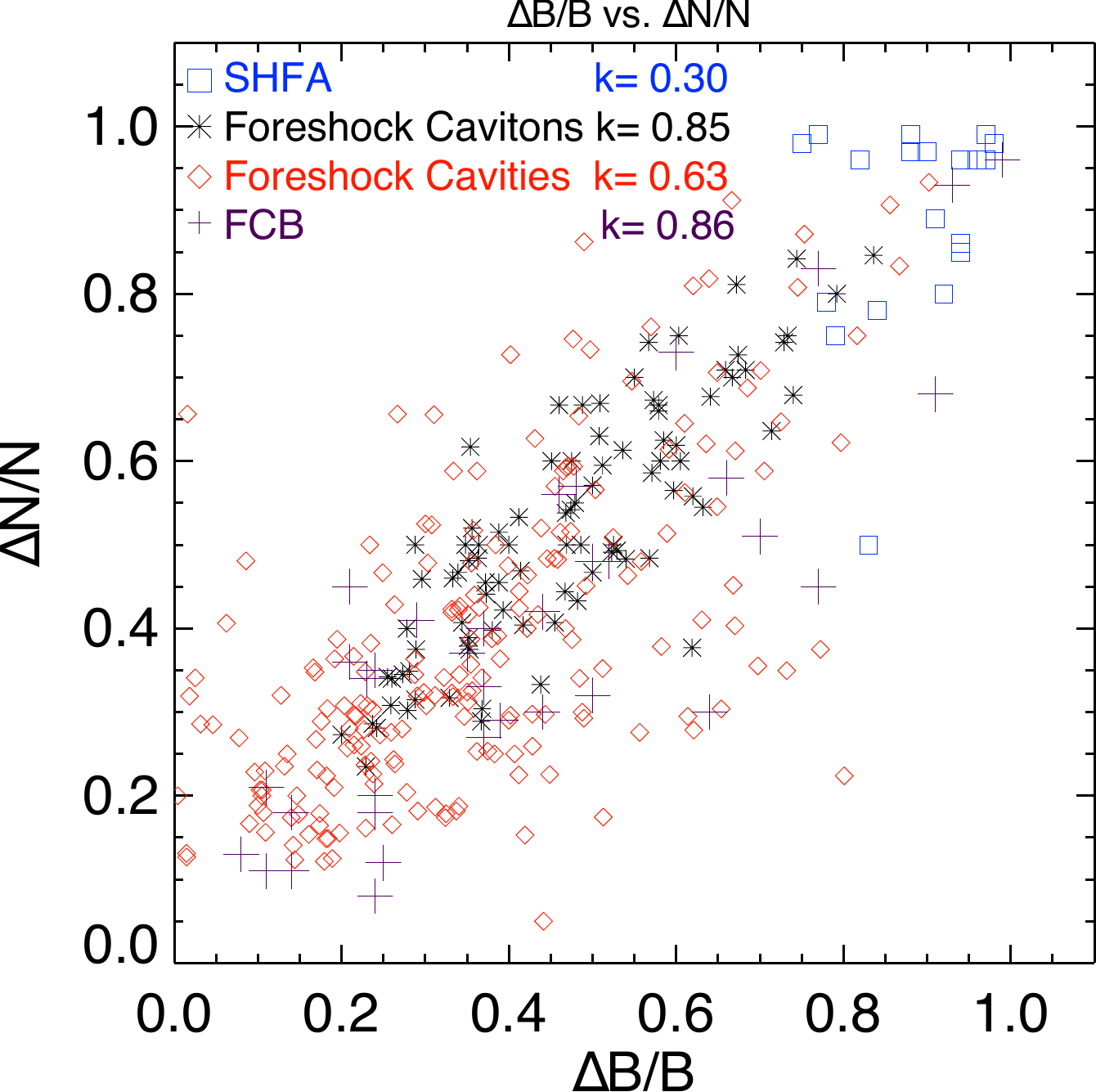}
\caption{Scatter plot of $\Delta$N/N versus $\Delta$B/B for the four types of upstream transient phenomena.}
\label{fig:dndb}
\end{figure*}

\begin{figure*}
\centering
\includegraphics[width = 0.9\textwidth]{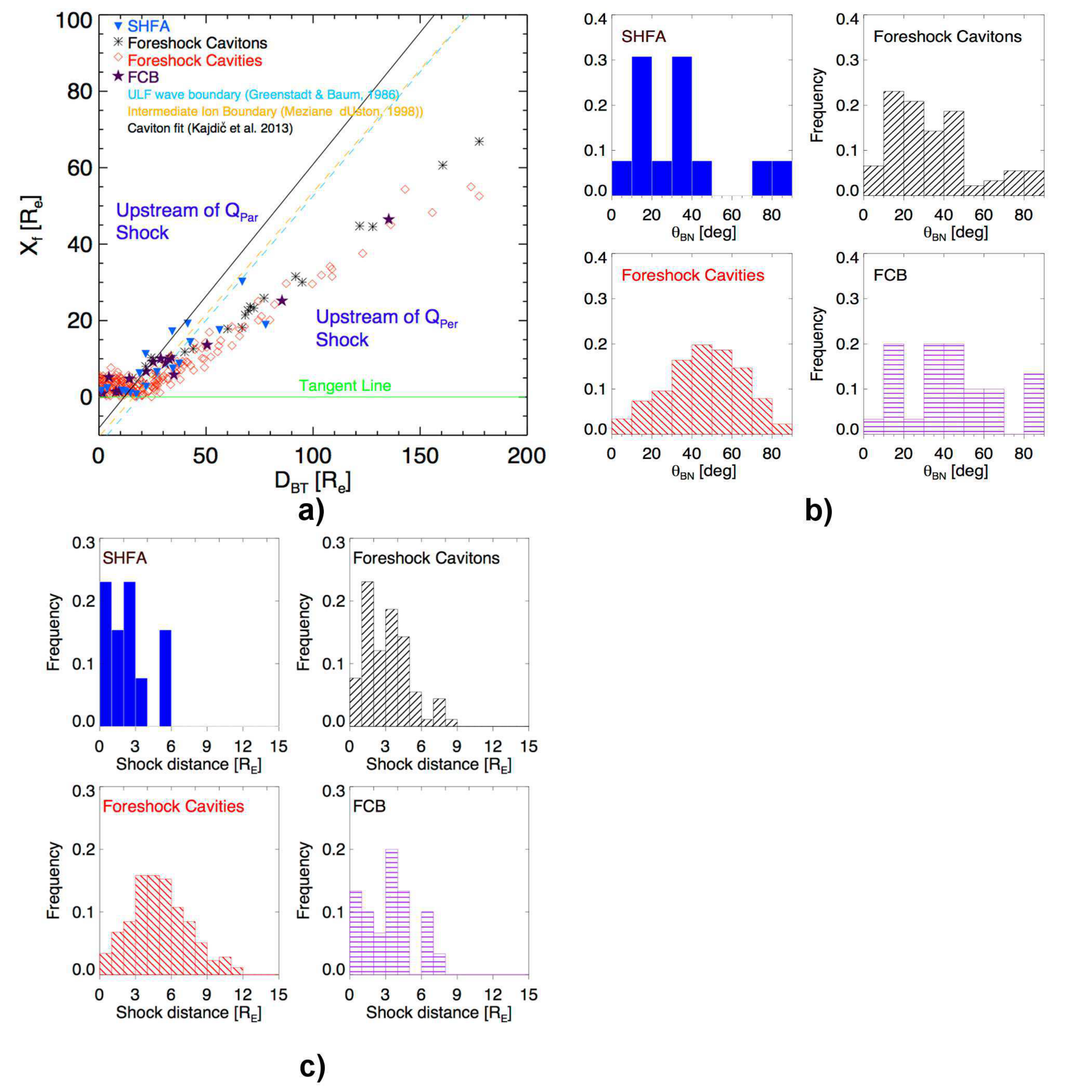}
\caption{a) Solar foreshock coordinates of the observed events. Black asterisks represent locations of foreshock cavitons, blue triangles those of the SHFAs, red diamonds of the foreshock cavities and purple stars of the FCBs. The horizontal green line represents a nominal tangent line. The dashed blue line is a fit to the ULF wave boundary by \citet{greenstadtbaum86} while the yellow dashed line represents a fit to ion intermediate boundary from \citet{mezianeduston98}. The black continuous line is a fit to caviton locations from \citet{kajdic13}. b) Distributions of the angles $\theta_{BN}$ of the portions of the bow-shock which different phenomena were magnetically connected to. c) Distance (along the X$_{GSE}$ axis) of the events to the model bow-shock.}
\label{fig:sfc}
\end{figure*}

\begin{figure*}
\centering
\includegraphics[width = .80\textwidth]{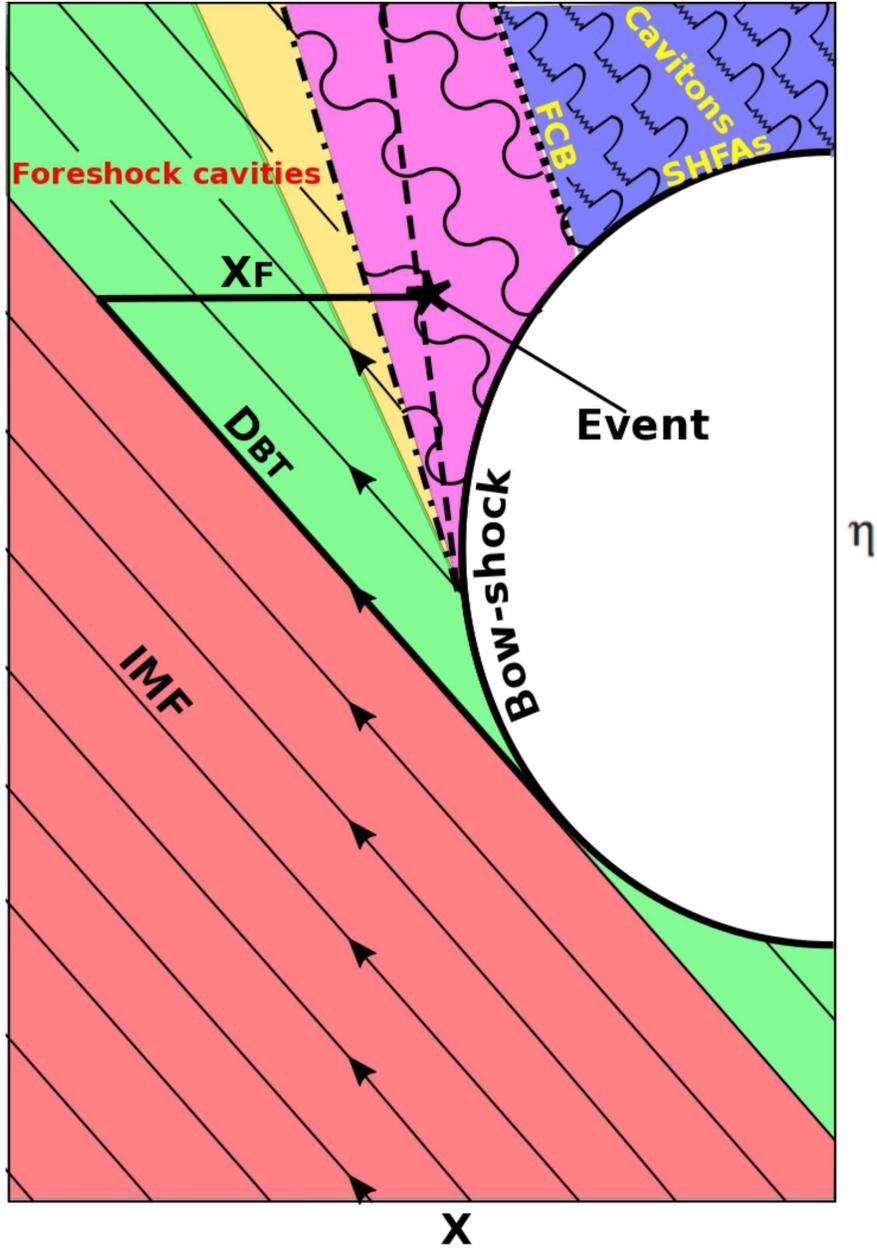}
\caption{Solar foreshock coordinates and different boundaries and regions upstream of the bow-shock. See text for details.}
\label{fig:sfcsystem}
\end{figure*}

\begin{figure*}
\centering
\includegraphics[width = 0.6\textwidth]{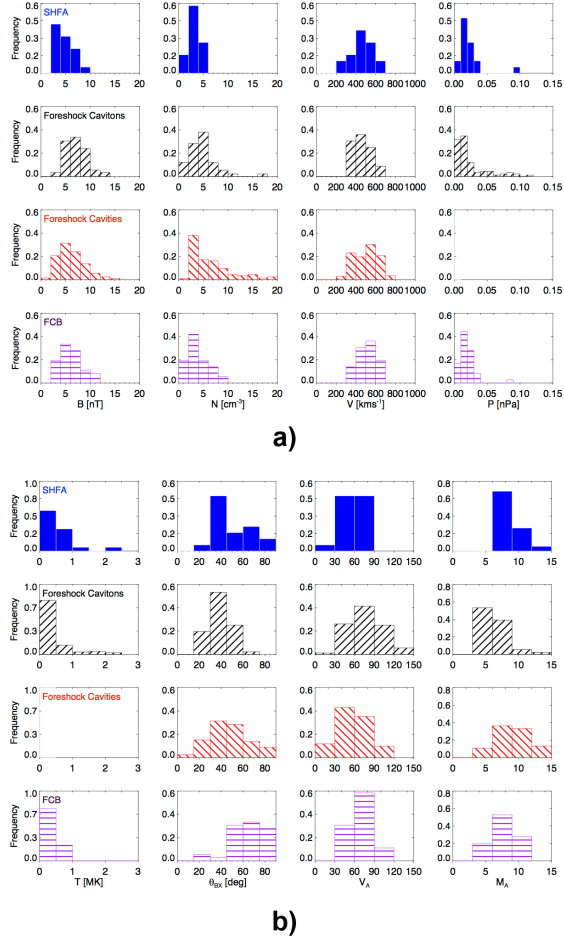}
\caption{a) Distributions of (from left to right) IMF magnitude, SW density, SW velocity and SW thermal pressure for times when (from top to bottom) SHFAs, foreshock cavitons, foreshock cavities and FCBs were observed. b) The same as a) but the distributions of the (from left to right) SW temperature, IMF cone angle $\theta_{BX}$, SW Alfv\'en velocity V$_A$ and Alfv\'enic Mach number M$_A$ are shown.}
\label{fig:swdist}
\end{figure*}

\end{document}